\documentclass[12pt, reqno]{amsart}

\newtheorem{theorem}{Theorem} 

\theoremstyle{definition}
\newtheorem{definition}{Definition}

\usepackage{setspace}
\usepackage{nomencl}
\usepackage{tabularx}
\makenomenclature
\usepackage{algorithm2e}

\usepackage{hyperref}
\usepackage[round]{natbib} 
\usepackage{epsfig}
\usepackage{rotating}
\usepackage{url}
\usepackage{mathtools}
\usepackage{amsfonts}
\usepackage{comment}
\usepackage{mathrsfs}
\usepackage{textcomp}
\usepackage{bm}
\usepackage{titletoc}
\usepackage[round]{natbib} 
\usepackage[OT1]{fontenc}
\usepackage{graphicx,epsfig}
\usepackage{epstopdf}
\usepackage{amscd}
\usepackage{amsmath,latexsym,amssymb,amsthm}
\usepackage{hyperref}
\hypersetup{colorlinks=true,citecolor=blue,linktocpage=true}
\newpage

\begin{document}

\title[Sampling from curved torus]{Sampling from the surface of a  curved torus:\\ A new genesis}


\maketitle
\begin{center}
\author{Buddhananda Banerjee}\\
\email{bbanerjee@maths.iitkgp.ac.in}\\
and\\
\author{Surojit Biswas}\\
\email{surojit23$@$iitkgp.ac.in}\\
\vspace{1cm}
\address{Department of Mathematics \\ Indian Institute of Technology Kharagpur, India-$721302$}\\

\end{center}

\begin{abstract}
  The distributions of toroidal data, often viewed as an extension of circular distributions, do not consider the intrinsic geometry of a curved torus. For the first time, \cite{diaconis2013sampling}[\textit{Diaconis, P., Holmes, S., \& Shahshahani, M. (2013). Sampling from a manifold. Advances in modern statistical theory and applications: a Festschrift in honor of Morris L. Eaton, 10, 102-125.}] introduce uniform distribution on the surface of a curved torus with respect to its surface area. But the suggested acceptance-rejection method of sampling from it rejects approximately half of the data. We propose a probabilistic transformation for sampling from the same distribution without losing data. In addition, we introduce a new genesis of random samples from some popular circular distributions using histogram-based acceptance-rejection sampling that uses a very thin envelope. The idea leads to generalizing for sampling from distributions on the surface of a curved torus with a high acceptance rate.Apart from reducing computational cost in the inferential study of different toroidal distributions,  uniform sampling from the surface of a curve torus will be helpful to understand any unknown distribution on it.


\end{abstract}
\newpage 

\tableofcontents{}

\newpage 





\tableofcontents{}

\nomenclature{\(\mathbb{R}^n\)}{n-dimensional real space}
\nomenclature{\(\mathbb{C}\)}{Space of complex numbers}
\nomenclature{\(\mathbb{N}\)}{Set of natural numbers}
\nomenclature{\(A^t\)}{Transpose of matrix A}

\nomenclature{\(\mathbb{S}^1\)}{Unit circle}
\nomenclature{\(\Omega\)}{Sample space}
\nomenclature{\(\mathcal{F}\)}{\sigma-Field}
\nomenclature{\(P\)}{Probability measure}
\nomenclature{\(\mathbb{P}\)}{Projection}
\nomenclature{\(\Omega\)}{Sample space}
\nomenclature{\(\mathbb{T}_a^{n}\)}{n-dimensional flat torus}
\nomenclature{\(\mathbb{T}_m^{n}\)}{n-dimensional general torus}
\nomenclature{\(\theta_1\)}{Horizontal angle}
\nomenclature{\(\theta_2\)}{Vertical angle}
\nomenclature{\(\mathcal{S}\)}{Parameter space}


\section{Introduction}

The analysis of the data from a surface or, in general, from a  manifold  depends on its genesis and representation. Such a process begins with sampling from a distribution on a manifold, and the data are used for performing statistical inferences such as parameter estimation, hypothesis testing, prediction, etc.  Drawing samples from the posterior distribution on constrained parameter spaces such as covariance matrices, testing goodness of fit for exponential families based on sufficient statistics, and producing data for assessing algorithms in topological statistics are examples of the instances  where sampling from a manifold is required.\\

Because of the availability of high dimensional dependent data,  the statistical inferences  on manifolds have been gaining more  attention now-a-days. \cite{bhattacharya2003large}, and \cite{pennec2006intrinsic}   have discussed the development of mean and variance estimators on manifolds. \cite{beran1979exponential}, \cite{Watson_1983}, and \cite{fisher1993statistical} have studied data on the  projective space and sphere.  An example, which involves atomic configurations where angles or inter-atomic distances are fixed, is commonly encountered in physics and chemistry; see \cite{fixman1974classical} and \cite{ciccotti1986molecular}.
Monte Carlo sampling from  manifolds becomes necessary in any of these above studies. Many well-established algorithms for sampling from the uniform distribution of homogeneous spaces and compact groups can be found in the literature. For example, if all the entries of an $n\times n $ random matrix follow standard normal and the QR decomposition is performed, then the Q part is uniformly distributed on the orthogonal group  \cite[see][]{eaton1983multivariate}. \cite{diaconis1986square} have studied the square roots of the uniform distribution on compact groups. There are available a few nice algorithms for sampling from the boundary of convex and compact sets in $\mathbb{R}^n$  \cite[see][]{belisle1993hit, boender1991shake, lalley1987asymptotically}. For a more general manifold, \cite{diaconis2013sampling} have developed an algorithm that makes it possible to sample from a probability distribution on a sub-manifold embedded  in $\mathbb{R}^n$. They have shown how their techniques can be used in a variety of contexts, such as the assessment of algorithms in topological statistics, conducting goodness of fit tests in exponential families, and performing Neyman's smoothness test.\\

\paragraph{\textbf{Example of astrophysics data:}}
     An active galactic nuclei (AGN) is a compact region at the core of a galaxy (known as an active galaxy) that is the consequence of gas falling onto a supermassive black hole (SBH). Hence, the gas forms a toroidal structure centered on the SBH due to the conservation of angular momentum.     
    An important characteristic of AGN is that it frequently exhibits a significant emission of infrared radiation, which is believed to come from a dusty torus surrounding the central black hole. In the context of the unified model, the different orientation of the torus relative to an observer from the earth explains the observational features of AGNs. The dusty torus around the SBH is treat as a major component in the understanding of AGN theories for quite some time. Still, the direct image on the appropriate physical scales was uncommon until    \cite{carilli2019imaging}  gives the first direct image of the thick torus in the active galactic nucleus (AGN) of one of the very powerful radio galaxy Cygnus A using NSF’s Karl G. Jansky Very Large Array (VLA).  So, uniform sampling of such toroidal data is essential to understand the fundamental behavior of AGNs.
    \\

 \paragraph{\textbf{Example of  biochemistry data:}}
 Protein folding is the physical process by which a chain of protein molecules changes into the final three-dimensional shape. The proper folding of a protein is essential for its function, as improper folding can result in a variety of diseases, such as neurodegenerative disorders. Identification and prediction of protein folding  is a major unsolved problem in biochemistry or computational biology. For a more comprehensive discussion, \cite[see][] {selkoe2003folding}. Due to the complex nature of the physical process involved in protein folding, there have been limited achievements in accurately predicting the final three-dimensional conformation of a protein from its sequence of amino acids. Improved comprehension of protein folding would undoubtedly result in clinical benefits, such as developing successful drug compounds for treating various diseases, including those mentioned earlier. Some portions of the structure of a protein may appear amorphous, necessitating the use of random models and probability distributions to characterize them precisely.
\cite{hingorani1998toroidal} and \cite{hingorani2000tale} discussed  that, in reality, a significant number of proteins with distinct evolutionary origins involved in DNA metabolism take on a toroidal form. The number of proteins that have this toroidal structure is relatively large. Due to the significance of the protein folding problem and the DNA-binding process, statisticians have taken interest in developing statistical models that can accurately depict these complex phenomena with toroidal data.\\

In this article, we are primarily interested in drawing random samples from different distributions on the surface of a curved torus. In a pioneering work, \cite{diaconis2013sampling}  proposed a method for obtaining uniform random samples from such a surface. By uniformity, they have considered the frequency of obtaining random samples  in proportion to the local area on the surface of the torus. Finally, they executed  the idea with acceptance-rejection sampling. However, the shortcoming of the suggested  sampling scheme  is that the rejection rate is remarkably high, approximately $50\%$, for one of the proposed marginal densities. We introduce  a probabilistic transformation that facilitates  drawing random samples from the target marginal  distribution without any rejection of the data. For understanding,  any unknown distribution on the surface of the torus essentially needs to have a uniform random sample from there. The proposed method will allow us to do the same with high efficiency. On top of that, we provide a genesis of random samples from some popular circular distributions using histogram-based acceptance-rejection  sampling. The idea is motivated by the upper-Riemann-sum of integration that  provides  a very thin envelope to the target density on a circle. The idea can be generalized  to draw random samples from different distributions on the surface of the curved torus incorporating its intrinsic geometry with a high acceptance rate. The proposed genesis of the data from the distributions on a circle or the  surface of a torus not only adds an advantage to the probabilistic  studies but also speed-up the simulation study of  the estimators and test statistics for inferential purposes.

The article is organized as follows.
 Section \ref{ch:background} begins with some fundamentals of geometric measure theory on the torus, followed by the intrinsic geometry of the same.
 In Section \ref{ch:EAU sec}, the Exact Area Uniform (EAU) sampling method has been introduced, which is an improved method to draw random samples from the uniform distribution on the surface of a curved torus over the proposed method by \cite{diaconis2013sampling}. In Section \ref{ch:HAR}, first, we introduced the idea of the Histogram-Acceptance-Rejection (HAR) sampling algorithm, which provides a very thin envelope leading to a very high acceptance rate for a finitely supported continuous distribution. We implement the HAR algorithm for von Mises distribution on a circle.  Then we extend the idea to generate samples from the surface of a curved torus. A detailed simulation has been reported in each of the sections. The concluding section is followed by the  necessary proofs provided in the Appendix-\ref{appendix}.

\section{Background}\label{ch:background}
A torus is a  geometric object representing two angular variables with respective radii. 
Only the angular part of it can be represented with a flat torus $([0,2\pi)\times [0,2\pi))$, but when the radii are involved, it is represented as the curved torus, see Eq. \ref{torus equation}. The curved torus is not homeomorphic to the flat torus because their topological properties differ. To analyze the data represented on the surfaces of a curved torus, it is essential to have a proper notion of probability distributions on it and statistical methodologies for the inference. The existing statistical techniques from the literature applicable  to the flat torus do not apply to the analysis of data on a curved torus because it does not take into account the topology and geometry of the surface. \\

Here, we focus on the  $2$-dimensional curved torus, a Riemannian manifold embedded in the $\mathbb{R}^{3}.$ In this report, we will use the term ``curved torus" for $2$-dimensional curved torus. The parameter space for the curved torus is $\mathcal{S}=\{ (\theta_1, \theta_2): 0\leq \theta_1, \theta_2<2\pi  \}$, and that can be represented in parametric equations as
 \begin{equation}
 \begin{aligned}
      x(\theta_1,\theta_2) &= (R+r\cos{\theta_2})\cos{\theta_1}\\
    y(\theta_1,\theta_2) &= (R+r\cos{\theta_2})\sin{\theta_1} \\
    z(\theta_1,\theta_2) &= r\sin{\theta_2},
 \end{aligned}
 \label{torus equation}
 \end{equation}
   where  $R,r$ are radii of the horizontal circle and vertical circle, respectively.  Let us consider some fundamentals of geometric measure theory from  \cite{federer2014geometric} for further developments.

\begin{definition}
A function $f:\mathbb{R}^m \xrightarrow{} \mathbb{R}^n$, is called
Lipschitzian function iff there exists a constant, $C>0,$ such that $|f(x)-f(y)|<C|x-y|$, whenever $x,y\in \mathbb{R}^m.$
\end{definition}

\begin{definition}
A set $E\subset \mathbb{R}^n,$ is called $m$-rectifiable iff it is a Lipschitzian image of some bounded subset of $\mathbb{R}^m.$
\end{definition}

\begin{definition}
   Let $A$ be any subset of $\mathbb{R}^n,$ the $m$-dimensional Hausdorff measure $\mathbb{H}_m(A)$ is defined by
\[ \mathbb{H}_m(A)=\lim_{\delta\to 0}~~\inf_{\substack {A\subseteq\cup S_i,\\ \text{diam}(S_i)\leq \delta}}  \sum \alpha_m \left(\dfrac{\text{diam}(S_i)}{2}\right)^m,\]
 
\end{definition}
where $\alpha_m=\dfrac{\Gamma(\frac{1}{2})^m}{\Gamma \left[(\frac{m}{2})+1\right]}$ which is the  volume of unit ball in $\mathbb{R}^n$, and the infimum is taken over all countable coverings $S_i$ of $A$ with $\text{diam}(S_i)= \sup \{|x-y|:x,y \in S_i  \}.$

The Hausdorff measure is an outer measure that acts as an area measure for subsets, and it satisfies the  countably additive property on the Borel sets of $\mathbb{R}^n$. \cite{federer2014geometric} shows that for a $m$-rectifiable set $A$, the covering can be restricted to the cubes or balls, and 
$$ \mathbb{H}_m(A)=\lim_{\epsilon\to0} \left[  \dfrac{\lambda_n (\{x: \text{dist}(x,A)<\epsilon\})}{\alpha_{n-m}\epsilon^{n-m}}
\right],$$ where $\lambda_n (dx)$ is the Lebesgue measure on the sets of $\mathbb{R}^n.$ 
The area formula is addressed in the following theorem \cite[see][] {federer2014geometric} is the natural extension of the concept of change of variables from differential calculus.

\begin{theorem}
Let $f:\mathbb{R}^m \xrightarrow{} \mathbb{R}^n$, be a
Lipschitzian function, and $m \leq n$, then
\begin{equation}
    \int_{A} g(f(x))J_mf(x)\lambda_m dx=\int_{\mathbb{R}^n} g(y)N(f|A,y)\mathbb{H}_m dy
\end{equation}
whenever $g:\mathbb{R}^n \xrightarrow{} \mathbb{R}$ is Borel, $A$ is $\lambda_m$ measurable and $N(f|A,y)$ is the cardinality of the set $\{x\in A: f(x)=y \}.$
\label{gm thm}
\end{theorem}

  In this article, the function  $f$ is a parameterization of the torus; see Eq. \ref{torus equation}.  So, the function $f$ is one-one, the integral on the right-hand side is the surface integral of $g$ over the set $f(A)\in \mathbb{R}^n$, and the left-hand side implements the integral using the Jacobian and the Lebesgue measure on $\mathbb{R}^m$, where in particular $m=2 \mbox{~and~} n=3.$ It demonstrates that sampling from the normalized density $J_mf(x)$  on $\mathbb{R}^m$  and then the image of that via $f$ onto the curved torus gives a sample from the surface of the curved torus in terms of area measure.\

The Jacobian plays an essential role in computing area measures. Let $f:\mathbb{R}^m \xrightarrow{} \mathbb{R}^n$ be a function, we say $f$ is differentiable at a point $x\in \mathbb{R}^m $ if there exists a linear map $L$ from $\mathbb{R}^m$ to $\mathbb{R}^n$ such that $$\dfrac{|f(x+h)-f(x)|}{|h|}\to 0 \mbox{~as~} h\to 0.$$ The  linear map is denoted by $Df(x)$ when it exists, and it can be determined by the partial derivatives 
$$
D_{i}(x)=\lim_{h \to 0}\dfrac{\left[f(x_1,\cdots, x_i+h, \cdots,x_n)-f(x_1, \cdots,x_n)\right]}{h}.
$$  
In matrix notation, the derivative matrix is  $(Df(x))_{i,j}=D_if_j(x)$ for $1\leq j\leq n$, and $1\leq i\leq m.$ Let $f:\mathbb{R}^m \xrightarrow{} \mathbb{R}^n$ be a differentiable function  at a point $x\in \mathbb{R}^m,$ the $p$-dimensional jacobian matrix, $J_pf(x)$ can be found by taking the norm of the derivative matrix. The following methods can be used to get the jacobian matrix for a given rank of $Df(x)$.  When the rank of $Df(x)<p$, $J_pf(x)=0.$
     When the rank of $Df(x)=p$, $J_p^2f(x)$ is equal to the sum of the squares of the determinants of the $p \times p$ sub-matrices of $Df(x)$.
     Usually, $p = m$ or $n$, then $J_p^2f(x)$
equals the determinant of the $p \times p$ product of transpose of $Df(x)$, $D^Tf(x)$ and $Df(x)$.
 For $p = m = n$, $J_p^2f(x)$ is the absolute value of the determinant of $Df(x)$.
 
 \subsection{Intrinsic geometry of torus}
The parametric equation of $2$-dimensional torus in Eq.  \ref{torus equation} is the Lipschitz image of the set $\{ 
 (\theta_1,\theta_2):0<\theta_1,\theta_2<2\pi\}\subset \mathbb{R}^2.$ Clearly, the function $f(\theta_1,\theta_2)=\{  (R+r\cos{\theta_2})\cos{\theta_1}, (R+r\cos{\theta_2})\sin{\theta_1}, r\sin{\theta_2} \}$ is a differentiable function from $\mathbb{R}^2$ to $\mathbb{R}^3$. Now, the partial derivatives of $f$ with respect to $\theta_1$, and $\theta_2$ are 
$$ \dfrac{\partial f}{\partial \theta_1}=\{ -(R+r\cos{\theta_2})\sin{\theta_1}, (R+r\cos{\theta_2})\cos{\theta_1}, 0 \}, $$ and
$$\dfrac{\partial f}{\partial \theta_{2}}=\{  -r\sin{\theta_2} \cos{\theta_1}, -r\sin{\theta_2}\sin{\theta_1}, r\cos{\theta_{2}}  \},$$ respectively. Hence, the derivative matrix is

$$Df(\theta_{1},\theta_{2})=\begin{bmatrix}
 -(R+r\cos{\theta_2})\sin{\theta_1} &  -r\sin{\theta_2} \cos{\theta_1} \\
 (R+r\cos{\theta_2})\cos{\theta_1} & -r\sin{\theta_2}\sin{\theta_1}\\
  0& r\cos{\theta_2}
\end{bmatrix}.$$ Therefore the jacobian is 

\begin{equation}
   J_{2}^{2}f(x)= \text{det} \left[ D^Tf(x) \cdot Df(x)\right]= \text{det} \begin{bmatrix}
 (R+r\cos{\theta_{2}})^{2} & 0 \\
 0 & r^{2}
\end{bmatrix}=r^{2}(R+r\cos{\theta_{2}})^{2}
\label{jacobian}
\end{equation}

Using the above expression of the square of the area element, which is the determinant of the product of transpose of the derivative matrix and the derivative matrix itself, \cite{diaconis2013sampling} proposed to draw the samples  $(\theta_{1},\theta_{2})$ from the density function given  in Eq. \ref{decompose equation} to ensure the uniformity with respect to area measure on the surface of a curved torus
\begin{equation}
    g(\theta_{1},\theta_{2})=\dfrac{(1+(r/R) \cos{\theta_{2}})}{4\pi^{2}}=g_1(\theta_1)~g_2(\theta_2),
    \label{decompose equation}
\end{equation}

where 
\begin{equation}
g_1(\theta_1)=\frac{1}{2\pi}, ~ 0\leq \theta_1<2\pi,
    \label{uni dist}
\end{equation}

and

\begin{equation}
g_2(\theta_2)=\frac{1}{2\pi}\left[1+\frac{r}{R} \cos{\theta_{2}}\right], ~ 0\leq \theta_2<2\pi. 
    \label{cos dist}
\end{equation}

The cumulative distribution function for $\theta_2$ is
$$
G_2(\theta_2)=\frac{1}{2\pi}\left[\theta_2+\frac{r}{R} \sin{\theta_{2}}\right], ~ 0\leq \theta_2<2\pi.
$$
\cite{diaconis2013sampling} use the acceptance-rejection sampling method for  generating samples from the density $g_2(\theta_2)$, and the  algorithm  for the same is also provided in their paper.

\section{Exact area uniform (EAU)  sampling from torus}\label{ch:EAU sec}
In this section, we proposed a new method to generate random uniform samples from the surface of the curved torus, which is equivalent to drawing samples from the densities $g_1(\theta_1)$ given in equation Eq. \ref{uni dist}, and  $g_2(\theta_2)$ given in Eq. \ref{cos dist}. In particular, we consider $a=r/R\in(0,1]$ and propose the method using a probabilistic transformation to generate samples from $g_2(\theta_2)$, describe in the Theorem-\ref{EAU thm}. The following Algorithm-\ref{alg:algo eau} is the pseudo-code for the new proposed EAU sampling method.

\begin{theorem}
    Let $U$ follows  uniform distribution on $[0, 1]$, $X$ follows  uniform distribution on $[0, 2\pi]$, and $p(X)=\dfrac{1}{2}(1+a\cos X)$. Then the random variable $Y$ is defined by \\
 
 $Y = \left\{
        \begin{array}{ll}
            X   & \mbox{if}  \quad U < p(X) ,\quad X<\pi\\
            \pi-X & \mbox{if}\quad U > p(X) ,\quad X<\pi\\
            X & \mbox{if} \quad U < p(X) ,\quad X>\pi\\
            3\pi-X & \mbox{if} \quad U > p(X) ,\quad X>\pi
        \end{array}
    \right. $\\

 follows the CDF,  $G(y)=\dfrac{\left(y+a \sin{y}\right)}{2\pi}$ where $0<a\leq 1$, and $ 0<y<2\pi$.
\begin{proof}
The proof of this theorem can be found in Appendix-\ref{appendix}.
\end{proof}
\label{EAU thm}	
\end{theorem}

\SetKwComment{Comment}{/* }{ */}
\RestyleAlgo{ruled}
\begin{algorithm}
\caption{EAU sampling algorithm}\label{alg:algo eau}
\KwData{  $(W,X) \gets n \mbox{~pairs of random sample from~}U[0,2\pi].$}

$a \gets \mbox{~ A number between~} 0 \mbox{~and~}1$\;
$pct\gets \dfrac{1+a \cos{X}}{2} $\Comment*[r]{define the probability}
$Y \gets \mbox{~Making an array of zeros of size~} n $\;
\For{$i \gets 1 \mbox{~ to~} n$}{
$rp[i] \gets \mbox{~Draw a Bernoulli random sample from~}Bernoulli(pct[i])$\;
  \eIf{$X[i]<\pi$ }{
    $a_1\gets  (X[i]*rp[i])+(\pi-X[i])*(1-rp[i])$\;
  }{\If{$X[i]>\pi$}{
      $a_2\gets  (X[i]*rp[i])+(3\pi-X[i])*(1-rp[i])$\;
    }
  }
  $Y \gets a_1+a_2 $\Comment*[r]{The random samples follow the desired distribution}
}
\KwResult{$Y \mbox{~ follows~} F(y)=\dfrac{\left(y+a \sin{y}\right)}{2\pi} \mbox{~where~}  0<a\leq 1, 0\leq y<2\pi. ~~ (W,Y)  $ are samples from area uniform distributed on the curved torus.}

\end{algorithm}

\subsection{Simulation Analysis} \label{uniform simulation}
In this subsection, we report the results of  a thorough simulation study to compare the proposed EAU sampling as  Algorithm-\ref{alg:algo eau} with the existing area uniform rejection (AUR) sampling algorithm provided by \cite{diaconis2013sampling}. We consider different values of $a=\frac{r}{R}$ form $0.1$ to $1$ with equal gaps and compute the acceptance percentages for the sample size of $10000$ in both algorithms. The following Table-\ref{table:eau uni} shows that the EAU sampling outperforms AUR sampling to a large extent. \

\begin{figure}[!h]

\includegraphics[width=0.8\textwidth,height=0.54\textwidth]{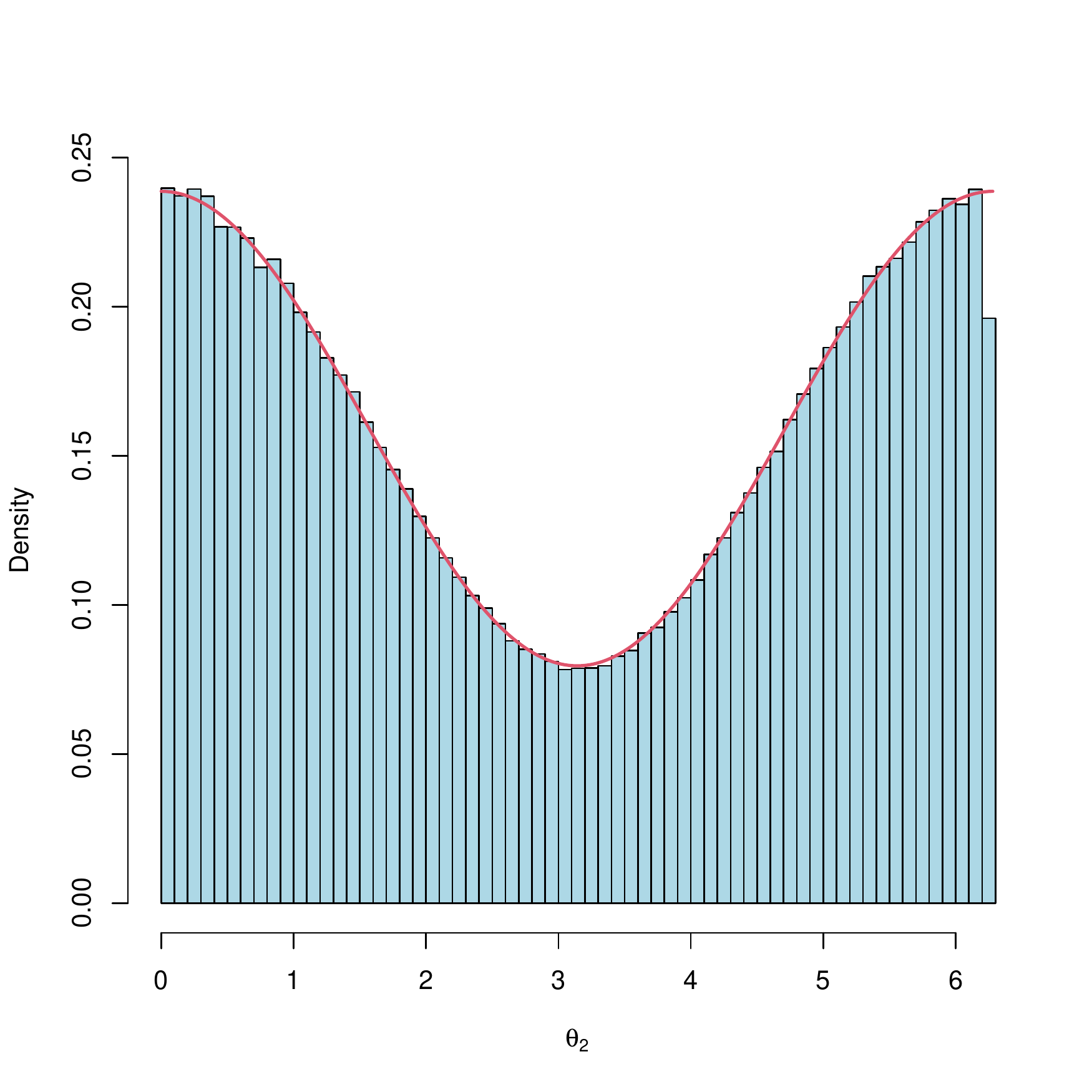}

	\caption{ Histogram of the sample from Theorem \ref{EAU thm} of the density $\frac{1}{2\pi}\left( 1+a\cos{\theta_{2}} \right).$ }
    \label{fig:density_hist}
\end{figure}

\begin{figure}[t]
	\includegraphics[width=0.7\textwidth, height= 0.55\textwidth]{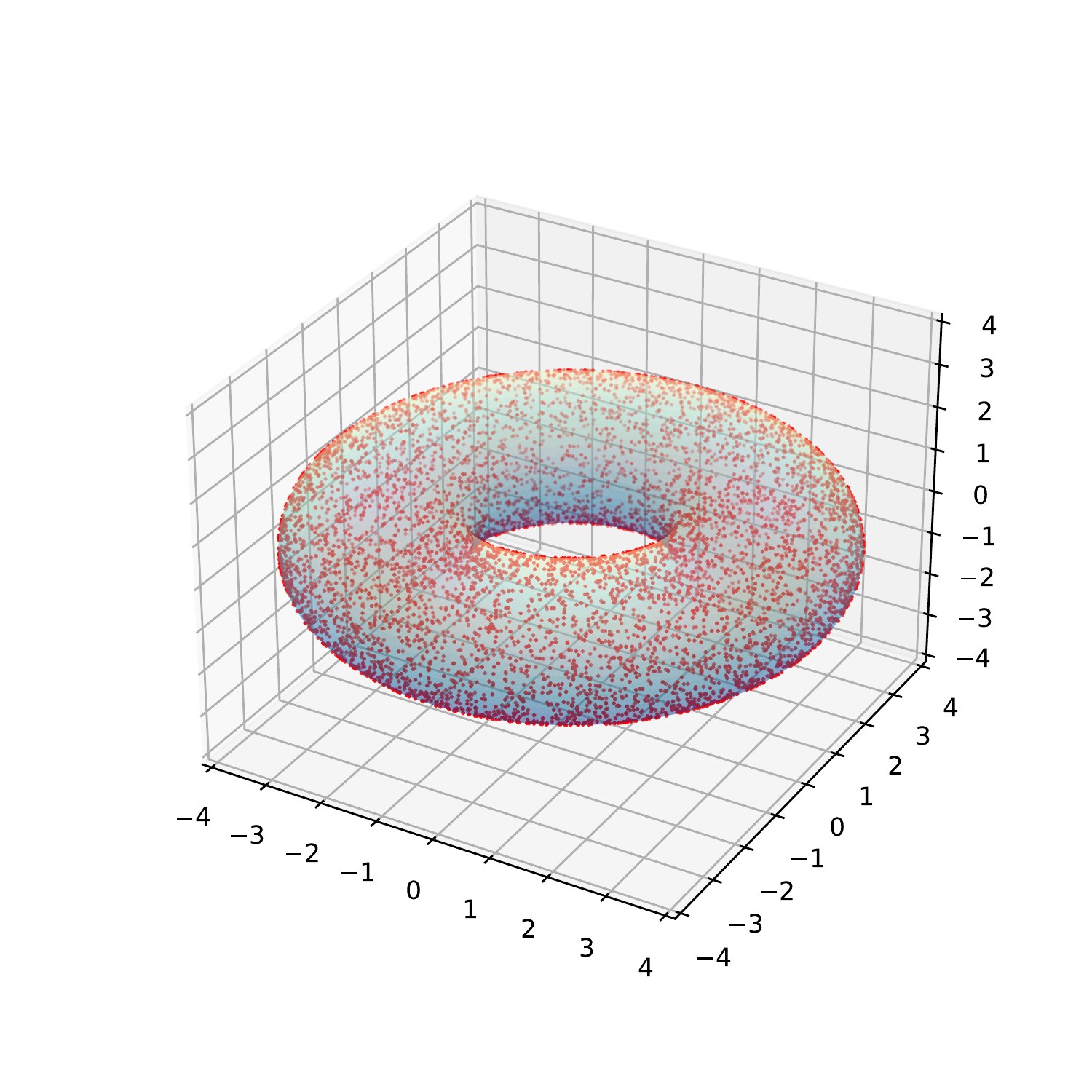}
	\caption{ Scatter plot of samples with EAU sampling method following uniform distribution  on the curved torus with respect to area measure. }
    \label{fig:uni area}
\end{figure}

\begin{figure}[!h]
	\includegraphics[width=0.7\textwidth, height= 0.55\textwidth]{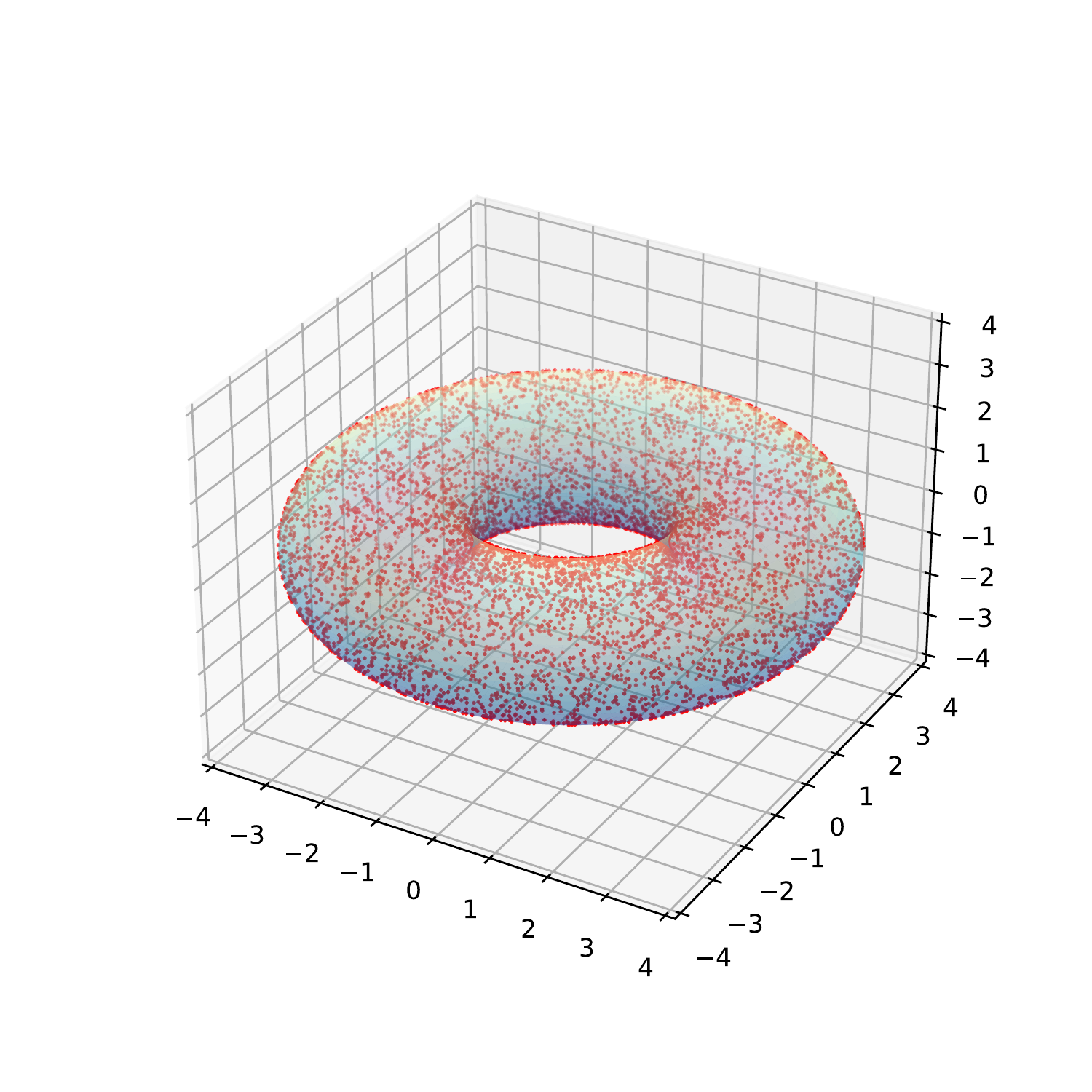}
	\caption{  Scatter plot of  points on torus when angular parameters are drawn uniformly from flat  torus}
    \label{fig:uni parameter}
\end{figure}

\begin{figure}[!h]
	\includegraphics[width=0.7\textwidth,height=0.35\textwidth]{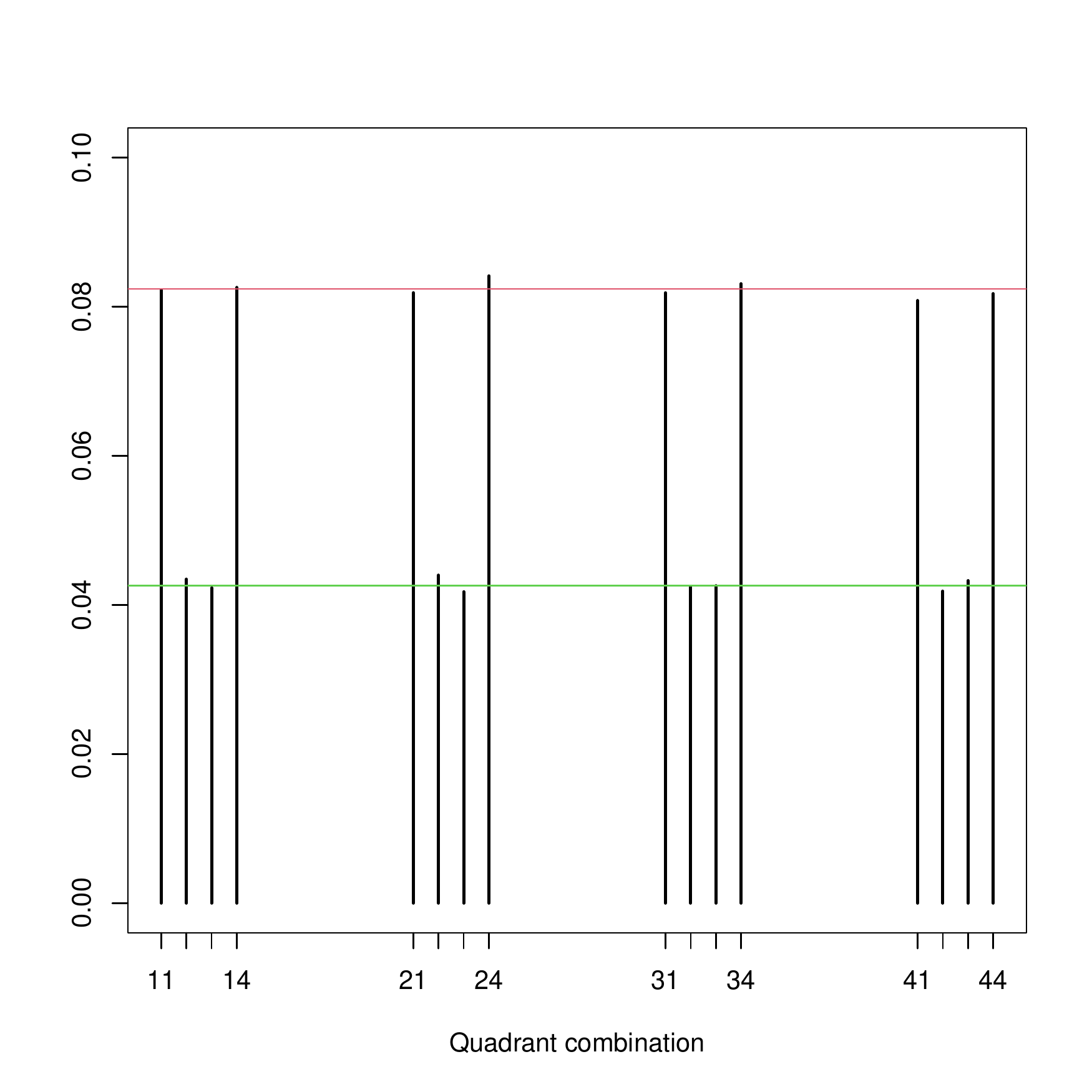}
	\caption{ Schematic of a contingency table of uniformly generated points on torus using EAU sampling method. The red and green lines are the area proportionate to the positive and negative curvatures, respectively.}
    \label{fig:con_algo}
\end{figure}

\begin{figure}[h!]
\includegraphics[width=0.7\textwidth,height=0.35\textwidth]{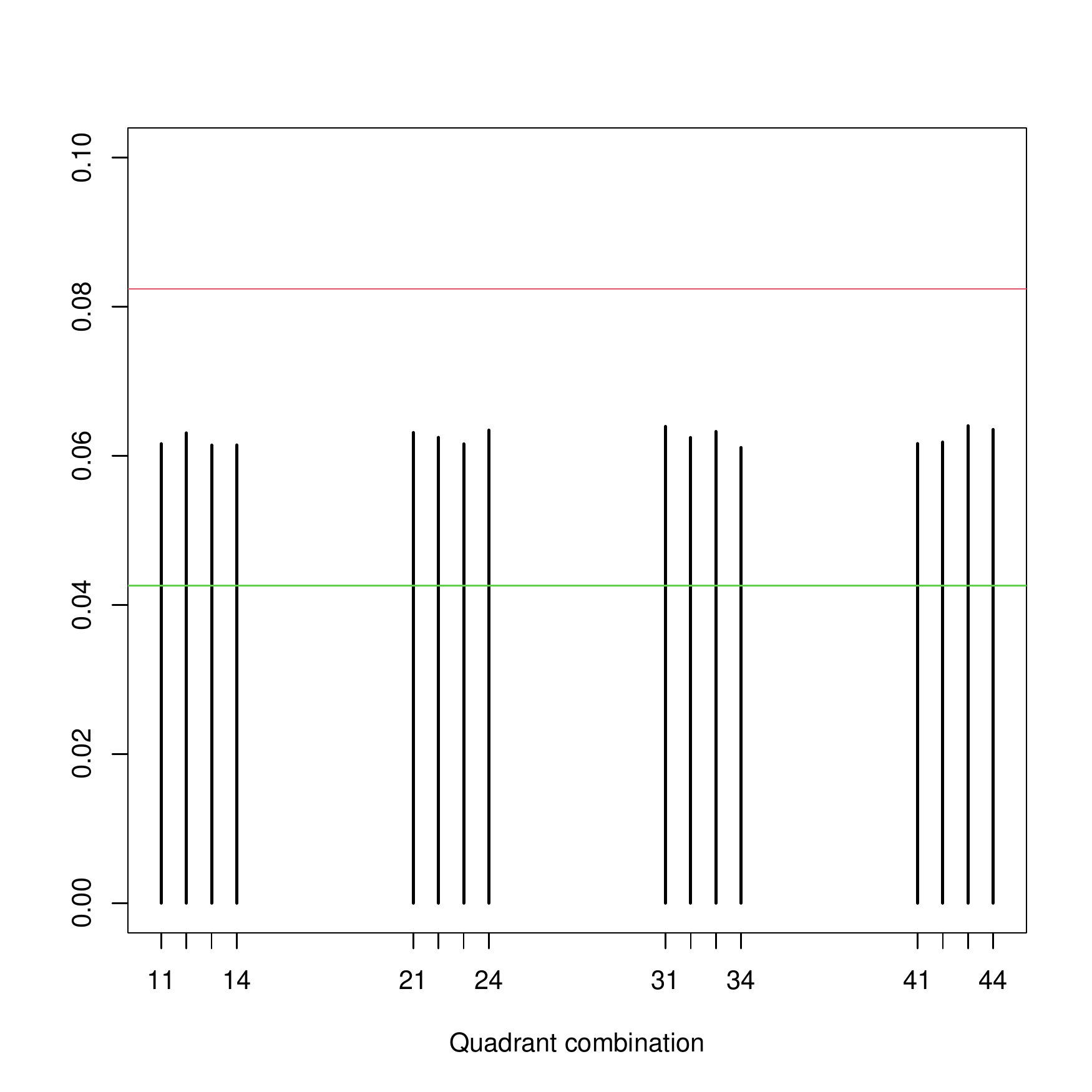}
	\caption{Schematic of a contingency table of points on the torus using the uniform distributions of angular parameters. The red and green lines are the area proportionate to the torus's positive curvature and negative curvature, respectively.}
    \label{fig:con_par}
\end{figure}

\begin{figure}[h!]
\includegraphics[width=0.7\textwidth,height=0.35\textwidth]{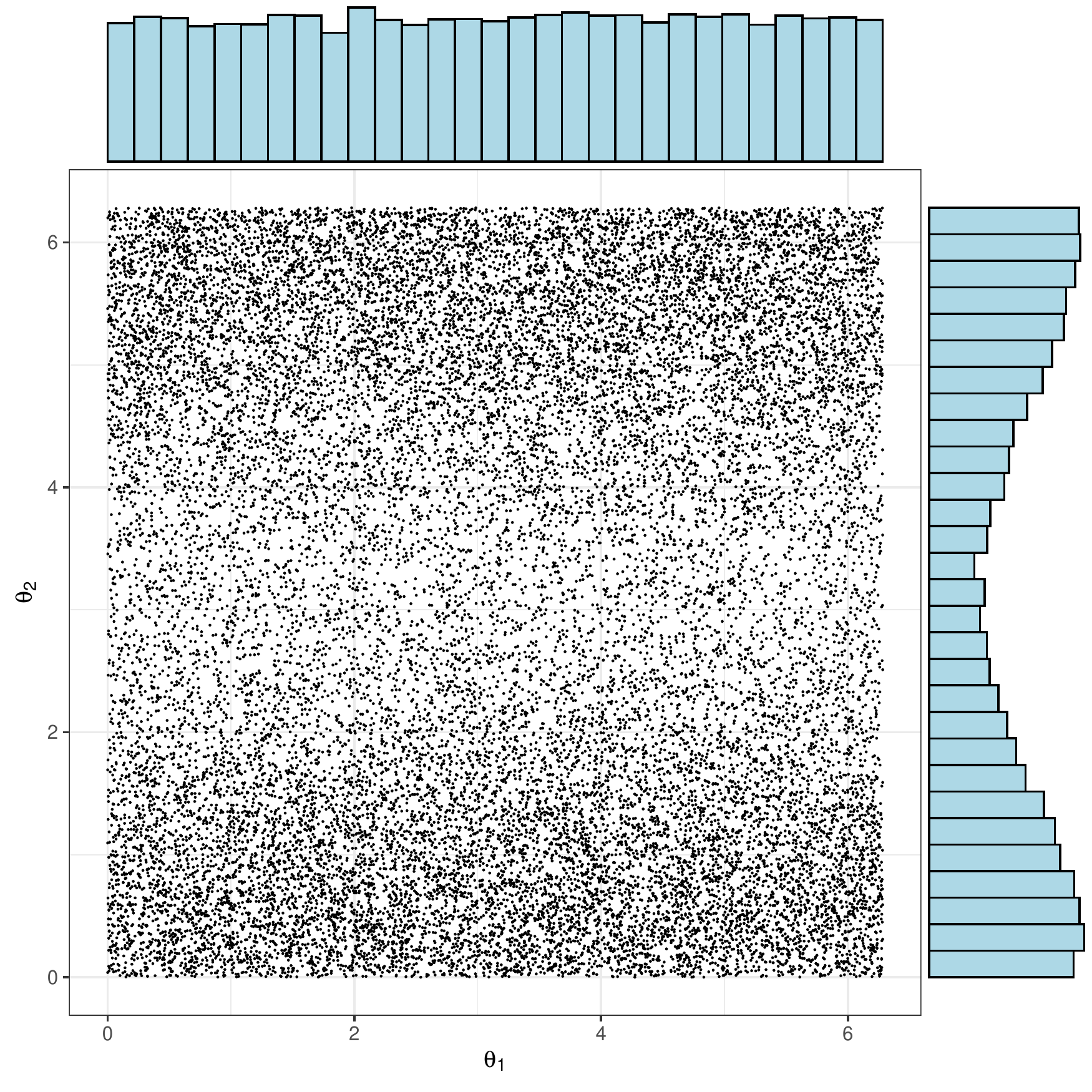}
	\caption{Plot of the marginal distributions of the data uniformly drawn from the surface of a curved torus with respect to
the area measure.}
    \label{fig:marginal eau}
\end{figure}

\begin{figure}[h!]
\includegraphics[width=0.7\textwidth,height=0.35\textwidth]{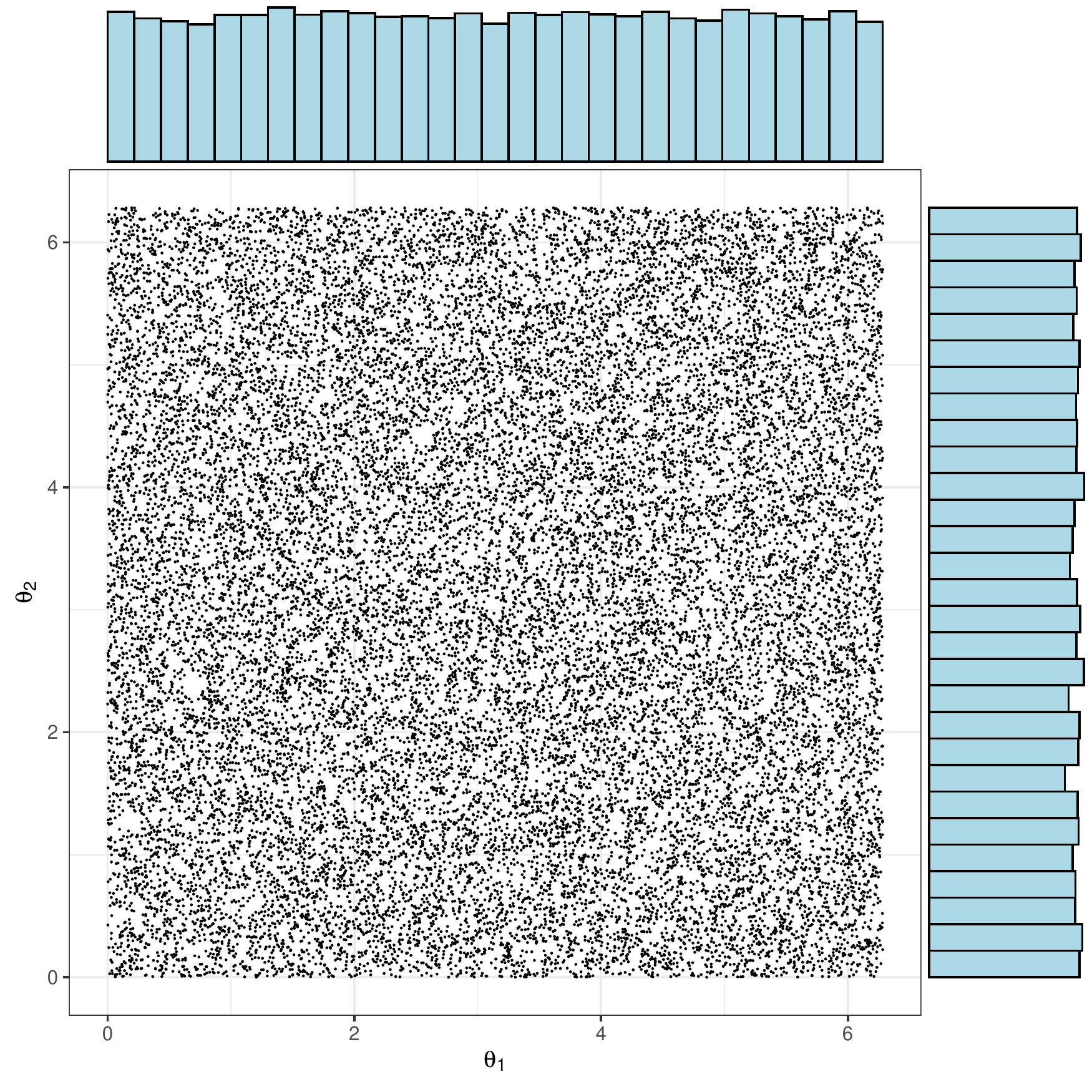}
	\caption{ Plot of the marginal distributions of the data uniformly drawn from its parameter space with respect to Lebesgue measure.}
    \label{fig:marginal flat}
\end{figure}

Figure-\ref{fig:density_hist} is the histogram of the sampled data from the marginal distribution of the vertical angle $\theta_2$ from the Eq. \ref{cos dist}. Figure-\ref{fig:uni area} is the scattered plot of the data  generated from the Algorithm-\ref{alg:algo eau} maintaining the uniform distribution using area measure on the surface of a curved torus  with $R=3, r=1.5$, and hence $a=0.5.$ Figure-\ref{fig:uni parameter} is the projection of the data uniformly generated from the flat torus to the surface of the curved torus. It is evident from the Figure-\ref{fig:uni parameter} that  
the region on the surface  with negative curvature has more density of data compared to that of Figure-\ref{fig:uni area}, which is drawn from EAU sampling.
 It might be hard to distinguish between  the scatter plots of Figure-\ref{fig:uni area}  and  Figure-\ref{fig:uni parameter}; a pair of bar diagrams of relative frequency can make the difference more prominent. We consider the quadrant combinations of horizontal and vertical circles as $Q_H\times Q_V$, with $Q_H=Q_V=\{1,2,3,4\}$, where each quadrant is of partition length $\pi/2.$ The bar diagram of Figure-\ref{fig:con_algo} represents the relative frequency of uniformly generated points on the torus using the EAU sampling method. Whereas the bar diagram of Figure-\ref{fig:con_par} represents the relative frequency of generated points from the flat torus with the uniform distributions of angular parameters. In both diagrams, the red and green lines represent the proportion of area  to the  quadrant combinations in positive and negative curvatures, respectively, to the total surface area of the curved torus. Although in  Figure-\ref{fig:con_algo}, the relative frequencies of quadrant combinations match with the respective proportions of the areas, in Figure-\ref{fig:con_par}, the same fail to do so. It can be also represented as the marginal plots of the data. Figure-\ref{fig:marginal eau} exhibits the marginal distribution of the data uniformly drawn from the surface of a curved torus with respect to the area measure, whereas Figure-\ref{fig:marginal flat} shows the same of the data  uniformly drawn from its parameter space which is  a flat torus. 
 Hence, the joint density function, $g(\theta_1, \theta_2)$, as in  Eq. \ref{decompose equation},   is the more natural notion of the uniform distribution on the surface area of a curved torus than on the flat torus, that is also advocated by \cite{diaconis2013sampling}.

 \begin{table}[!h]
\centering
\begin{tabular}{|l|c|c|c|c|c|c|c|c|c|c|}
\hline
\textbf{a}=$\frac{r}{R}$ & $0.1$ & $0.2$ & $0.3$ & $0.4$ & $0.5$ & $0.6$ & $0.7$ & $0.8$ & $0.9$ & $1$\\
\hline
\hline
\textbf{EAU} & $100 $ & $100$ & $100$ & $100$ & $100$ & $100$ & $100$ & $100$ & $100$ & $100$\\

\hline
\textbf{AUR} & $50.35 $ & $51.07 $  & $49.91 $ & $49.68 $ & $50.14 $ & $49.82 $  & $49.88 $  & $49.41 $  & $49.96 $ & $49.41 $\\
\hline
\end{tabular}
\vspace{.1cm}
\caption{Acceptance percentage comparison table  for uniform distribution on the curved torus.}
\label{table:eau uni}
\end{table}


\newpage
\section{Histogram-Acceptance-Rejection (HAR) Sampling}\label{ch:HAR}
In many scientific and technical applications, modeling angular or circular data is essential. For theoretical and practical studies simulating from angular distribution is also very important. Acceptance-rejection sampling is one of the most applicable approaches for sampling from angular distributions, such as von Mises distribution, cardioid, and Kato-Jones distributions, etc. The main challenge in the  acceptance-rejection sampling method is to find a suitable envelope on which the acceptance probability depends a lot. 
 While doing a simulation in large numbers, it is computationally expensive if the method  rejects a large proportion of samples just because of the choice of the envelope.
Here, we proposed a new methodology, say \textit{Histogram-Acceptance-Rejection } (HAR) sampling, for enveloping the target distribution motivated by the construction of the upper Riemann sum  so that the number of the rejected sample can be substantially reduced. The HAR  sampling method can be implementable for a probability density function which is  Riemann integrable on a bounded interval. In particular, for this article, we focus on some popular circular distributions, which naturally satisfy the above conditions. In  Riemann integration, the upper Riemann sum, which is the area under the  dominating step function of a non-negative integrated, consists of more area  than that of the actual integrand. But the dominating step function can be normalized with  up-to-the-total area one such that it can be considered as the legitimate probability density function that eventually can be used as a proposed density for a pre-specified target distribution. In the following subsection, we describe the methodology in detail.

\subsection{Proposed method for HAR Sampling}
Let $U$ follow uniform distribution on $[0,1]$.
Assume that $f(x)$ and $p(x)$ be the target and proposed probability  density functions, respectively, with the common finite  support $[a,b]$. Consider a  partitions of the interval $[a,b]$ as: $a=a_0<a_1<a_2<\cdots<a_{i-1}<a_i<\cdots<a_{k-1}<a_k=b$.
Now, for $i=1,\cdots, k$, let 
$ A_{i} = [a_{i-1},a_i],$ $ B = a_{i}-a_{i-1}=\frac{(b-a)}{k}, $ and $P(A_i) = \displaystyle \int_{A_i}f(y) dy $.
 Let, for the $i^{th}$ cell $H_i=\max\limits_{x\in A_i}
 f(x)$, $Y_i=a_{i-1}+BU$. So, the proposed density for the entire support is 
 \begin{equation}
     p(y)= \sum_{i=1}^{k} \dfrac{H_{i} \textbf{I}_{A_i}(y)}{\displaystyle B\sum_{i=1}^{k} H_i}
     \label{proposed density}
 \end{equation}
 satisfying the condition that $p(y \mid y\in A_i) =\frac{1}{B}$.
 We choose the number, $M_i=\frac{BH_i}{P(A_i)}$ for the $i^{th}$ cell such that $M_i\geq \max\limits_{y\in A_i}\frac{f(y\mid y\in A_i)}{p(y \mid y\in A_i)}.$
 Consider the cumulative distribution function (CDF) of $X$ for the $i^{th}$ cell as:

 \begingroup
\allowdisplaybreaks
\begin{align}
P(X\leq x)&=P(Y_i\leq x \,| \, Y_i  \hspace{.2cm}\text{accepted}) \nonumber \\
&=\dfrac{P\left(Y_i\leq x, \hspace{.2cm} U<\dfrac{f(Y_i)/P(A_i)}{M_i(\frac{1}
{B})}\right) }{P\left(U<\dfrac{f(Y_i)/P(A_i)}{M_i(\frac{1}
{B})}\right) }
  \label{rejection sampling1i}  
\end{align}%
\endgroup
Now considering the numerator
\begin{equation}
    \begin{aligned}
P\left(Y_i\leq x, \hspace{.2cm} U<\dfrac{f(Y_i)/P(A_i)}{M_i(\frac{1}
{B})}\right) &=\int P\left(Y_i\leq x, \hspace{.2cm} U<\dfrac{f(Y_i)/P(A_i)}{M_i(\frac{1}
{B})} \middle | Y_i=y \right)\left( \frac{1}{B}\right) dy  \\
&=\int \textbf{I}_{(y\leq x)} P\left( U<\dfrac{f(y)/P(A_i)}{M_i(\frac{1}
{B})}\right) \left( \frac{1}{B}\right) dy.\\
\end{aligned}
\label{rejection sampling1i1}
\end{equation}

Similarly,
\begin{equation}
\begin{aligned}
P\left(U<\dfrac{f(Y_i)/P(A_i)}{M_i(\frac{1}
{B})}\right) &=\int_{A_i} \dfrac{f(y)/P(A_i)}{M_i(\frac{1}
{B})}\left( \frac{1}{B}\right) dy \\
\end{aligned}
\label{rejection sampling1i2}
\end{equation}

Using the results of Eq. \ref{rejection sampling1i1}, and Eq. \ref{rejection sampling1i2} in Eq. \ref{rejection sampling1i} we get

\begin{equation}
    \begin{aligned}
P(Y_i\leq x\,| Y_i  \hspace{.2cm}\text{accepted})
&=\dfrac{\displaystyle \int \textbf{I}_{(y\leq x)} P\left( U<\dfrac{f(y)/P(A_i)}{M_i(\frac{1}
{B})}\right) \left( \frac{1}{B}\right) dy}{\displaystyle 
 \int_{A_i} \dfrac{f(y)/P(A_i)}{M_i(\frac{1}
{B})}\left( \frac{1}{B}\right) dy.}\\
\end{aligned}
\label{rejection sampling1ic}  
\end{equation}

The cumulative distribution function of $X$ for the entire range is given by 
\begin{equation}
    P(X\leq x)= \int_{a}^{x} f(t)\,\,dt
    =  \left(\frac{1}{M} \int_{a}^{x} f(t)\,\,dt \right) \Bigg/ \left(\frac{1}{M}\right),
\label{rejection sampling2}  
\end{equation} 

where we can choose $M$ in such a way that

\[        M \geq \displaystyle \max_{x}\, \frac{f(x)}{p(x)} 
   =\displaystyle \max_{i} \biggl\{  \max_{x\in A_{i}} ~ \frac{f(x)}{p(x)}  \biggl\}
   = \displaystyle \max_{1\leq i \leq k}  \left[ H_{i} \Bigg/ \left( \frac{H_{i}}{B~\sum_{i=1}^{k}H_{i}} \right) \right]  = \displaystyle B\, \sum_{i=1}^{k}H_{i}. \]

Note that
\begingroup
\allowdisplaybreaks
\begin{align}
 \frac{1}{M} \int_{a}^{x} f(t)\,\,dt 
     &=  \sum_{i=1}^{k} \left[ \frac{1}{M} \int_{A_i} \textbf{I}_{(t\leq x)}\,\, \dfrac{f(t)}{p(t)} \,\, p(t) \,\, dt \right] \nonumber\\
     &=  \sum_{i=1}^{k} \left(\frac{P(A_{i})}{M} \right) \left[ 
     \dfrac{\displaystyle \int_{A_i} \textbf{I}_{(t\leq x)}\,\, P\left( U<\dfrac{f(t)/P(A_i)}{M_i(\frac{1}
{B})}\right) \,\, \left(\frac{1}{B}\right) \,\, dt }{\displaystyle \frac{1}{1/M_{i}} }
     \right],
\label{rejection sampling21}
\end{align}%
\endgroup
for details see Appendix-\ref{appendix}. Hence, using the Eq. \ref{rejection sampling1ic} and Eq. \ref{rejection sampling21}  in Eq. \ref{rejection sampling2} we get
\begin{equation}
      \dfrac{\displaystyle \frac{1}{M} \int_{a}^{x} f(t)\,\,dt }{\displaystyle \frac{1}{M}}  =  \sum_{i=1}^{k}  P(A_{i})  \left[ 
     \dfrac{\displaystyle \int_{A_i} \textbf{I}_{(t\leq x)}\,\, P\left( U<\dfrac{f(t)/P(A_i)}{M_i(\frac{1}
{B})}\right) \,\, \left(\frac{1}{B}\right) \,\, dt }{\displaystyle 
 \int_{A_i} \dfrac{f(t)/P(A_i)}{M_i(\frac{1}
{B})}\left( \frac{1}{B}\right) dt.}
     \right].  
\end{equation}
As a consequence
$
  \displaystyle  \int_{a}^{x} f(t)\,\,dt =\sum_{i=1}^{k} P(A_{i}) \left[ P(Y_i\leq x \,| \,Y_i  \hspace{.2cm}\text{accepted})\right],
$ which can be implemented in 
 the following Algorithm-\ref{alg:algo HAR} that is the pseudo-code of the proposed HAR sampling method.

\SetKwComment{Comment}{/* }{ */}
\RestyleAlgo{ruled}
\begin{algorithm}
\caption{HAR sampling algorithm}\label{alg:algo HAR}
\KwData{ Target probability density function $f(x)$ with support $[a,b]$.}
$n \gets$ Number of random samples to be generated\;
$np \gets$ Number of partitions with equal length\;
$pt\gets$  A sequence in $[a,b]$ with length $(np+1)$ \;
$H\gets f(pt)$ \Comment*[r]{ Heights of the probability density function at pt}
$H_m\gets$ A sequence of the maximum heights of each partition \;
$bl\gets \frac{b-a}{np}$ \Comment*[r]{ Bin length } 
$p_m \gets \displaystyle \frac{H_m}{\sum H_m}$\Comment*[r]{ Probability vector}
  $count=0$\;
  $y \gets \mbox{~Initialize a vector of size zero} $\;
  \While{$count \leq n$}{
  $u\gets$ Draw a number random  from $U[0,1]$ \;
$ml \gets $ Draw a random sample from multinomial with probability vector $p_m$ \;
$x \gets pt[ml]+u*bl$\;
$px\gets \frac{f(x)}{H_m[ml]}$\;
$rp \gets$ A random number from $Bernoulli(px)$\;
\If{$rp=1$}{
    $count=count+1$\;
    $y[count]\gets$  $x$\;
  }

}
\KwResult{Samples from the probability density function $f(x)$ with support $[a,b]$.}
\end{algorithm}

\subsection{Simulation from von Mises distribution}
The probability distribution function 
\begin{equation}
    f(\theta)=\frac{e^{\kappa\cos(\theta-\mu)}}{2\pi I_{0}(\kappa)},
\label{von mises}
\end{equation}
where $0\leq \theta<2\pi$, $0\leq \mu<2\pi$, $\kappa>0$, and $ I_{0}(\kappa)$ is the modified Bessel function with order zero evaluated at $\kappa,$ defines the von Mises distribution. It is a widely applicable probability distribution for modeling angular or circular data. From Eq. \ref{von mises}, it is evident that the probability distribution with concentration parameter $\kappa$ is a continuous and symmetric  distribution with respect to the mean direction $\mu$. 
Sampling from the von Mises distribution is challenging because its cumulative distribution function (CDF) does not have any closed form, as discussed in \cite{mardia2000directional}. Therefore, conventional sampling methods such as inverse transform sampling cannot be used for this distribution.\

The sampling procedure proposed by \cite{best1979efficient} for the von Mises distribution is a well-known method that uses the conventional rejection sampling technique with the wrapped Cauchy distribution as an envelope, which we refer to as vMBFR sampling. However, the vMBFR sampling method   has a high rejection rate depending on the parameter. On the contrary, the proposed HAR sampling method has a much lower rejection rate. \

\begin{figure}[t]
	\includegraphics[width=0.9\textwidth,height=0.55\textwidth]{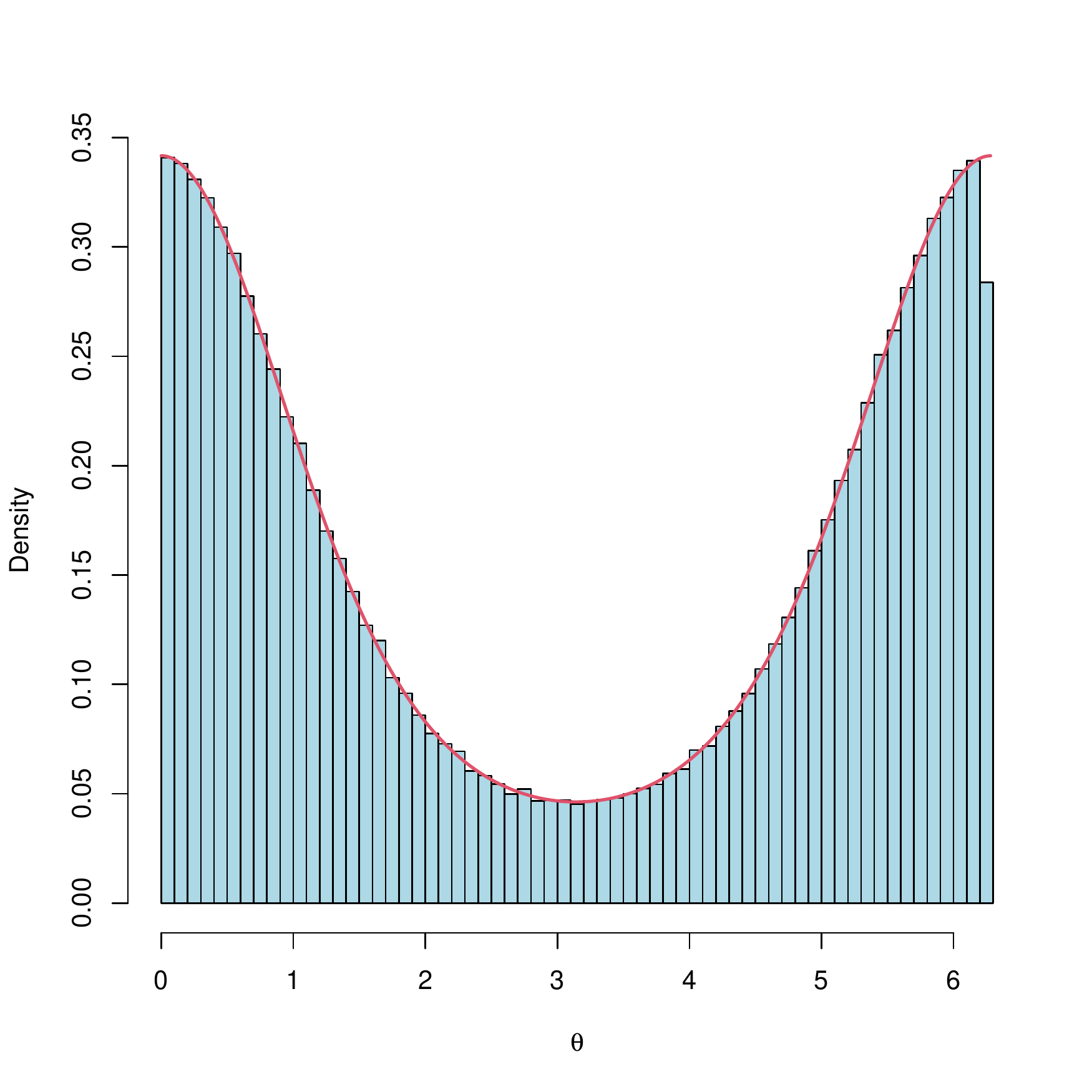}
	\caption{ Histogram of the sample data from von Mises distribution generated by HAR sampling method. }
    \label{fig:vm circle}
\end{figure}

In this study, we conducted a simulation with a sample size of $n=50000$ to compare the acceptance percentage of sample size between  HAR and vMBFR algorithms.  Table-\ref{table:vm HAR1} and Table-\ref{table:vm HAR2}  present the acceptance percentage for different values of the concentration parameter $\kappa$ with a mean direction parameter $\mu=0$. The proposed HAR sampling method outperforms the existing vMBFR sampling method in terms of the acceptance percentage of sample size. Figure-\ref{fig:vm circle} displays the histogram of the sampled points from the von Mises distribution using the HAR sampling technique,

 \begin{table}[h!]
\centering
\begin{tabular}{|l|c|c|c|c|c|c|c|c|c|c|}
\hline
$\kappa$ & $0.1$ & $0.2$ & $0.3$ & $0.4$ & $0.5$ & $0.6$ & $0.7$ & $0.8$ & $0.9$ & $1$\\
\hline
\hline
\textbf{HAR } & $99.96$ & $99.92$ & $99.87$ & $99.85$ & $99.81$ & $99.77$ & $99.72$ & $99.71$ & $99.67$ & $99.65$\\

\hline
\textbf{vMBFR } & $99.76$ & $99.06$  & $97.90$ & $96.67$ & $95.04$   & $93.23$  & $91.88$  & $89.88$ & $88.12$ & $86.94 $\\
\hline
\end{tabular}
\vspace{.1cm}
\caption{Acceptance percentage comparison table for von Mises distribution.}
\label{table:vm HAR1}

\centering
\scalebox{1}{
\begin{tabular}{|l|c|c|c|c|c|c|c|c|c|c|}
\hline
$\kappa$ & $2$ & $3$ & $4$ & $5$ & $10$ & $20$ &$40$ & $60$  &$80$ &$100$ \\
\hline
\hline
\textbf{HAR } & $99.48$ & $99.21$ & $99.02$ & $98.91$ & $98.462$ & $97.76$ &$96.96$ & $96.31$ &$96.76$ & $95.15$ \\

\hline
\textbf{vMBFR } & $76.95$ & $72.37$  & $69.96$ & $69.46$ & $67.46$   & $66.64$  & $66.43$  & $65.96$ &$65.94$ &$65.69$\\
\hline
\end{tabular}}
\vspace{.1cm}
\caption{Acceptance percentage comparison table for von Mises distribution.}
\label{table:vm HAR2}
\end{table}

In a similar way, the HAR sampling method can be implemented for cardioid, Kato-Jones distributions, etc. apart from the von Mises distribution for a greater acceptance rate.

\section{Generalization to sampling from the surface of torus} \label{torus sampling}
Although several  distributions on the flat torus have been introduced and studied by many researchers, only the distribution on the surface of the curved torus has been investigated by \cite{diaconis2013sampling}. They have specially studied uniform distribution on torus given in Eq. 
\ref{torus equation} as an example of a manifold.
In this section, we introduced a generalization for some popular distributions on the surface of a curved torus as an extension of circular distribution in  higher dimensions.

Applying the principles from differential geometry discussed in Subsection-\ref{ch:background}, from Eq. \ref{jacobian} we can determine the  area element of the torus as $dA=r(R+r\cos{\theta_{2}}).$ Hence, it is immediate that the area of the torus is given by
\begin{equation}
    A = \int_{0}^{2\pi} \int_{0}^{2\pi}  r(R+r\cos{\theta_{2}})\,\,d\theta_{1}\,d\theta_{2}=4\pi^2rR
\label{area torus}
\end{equation}
Now, let us consider a joint probability density function, $h(\theta_1,\,\theta_2)$ of $\theta_1$ and $\theta_2$, and from the Eq. \ref{area torus} obtain the identity

\begin{eqnarray}
 4\pi^2rR &=& \int_{0}^{2\pi} \int_{0}^{2\pi} h(\theta_1,\,\theta_2) (2\pi r)(2\pi R)\,d\theta_{1}\,d\theta_{2} \nonumber\\
     &=& \frac{1}{C}\int_{0}^{2\pi} \int_{0}^{2\pi} h(\theta_1,\,\theta_2) \left(1+\frac{r}{R}\cos\theta_2 \right)
     (2\pi r)(2\pi R)\,d\theta_{1}\,d\theta_{2},\nonumber
\end{eqnarray}
where $C$ is a normalizing constant, implying

\begin{eqnarray}
      1&=& \frac{1}{C}\int_{0}^{2\pi} \int_{0}^{2\pi} h(\theta_1,\,\theta_2) \left(1+\frac{r}{R}\cos\theta_2 \right)
     \,d\theta_{1}\,d\theta_{2}\nonumber\\
      &=&   \frac{1}{C}  \int_{0}^{2\pi} \int_{0}^{2\pi} h_{1}(\theta_1 \,|\, \theta_2)\, \left[\,h_{2}(\theta_2)\left(1+\frac{r}{R}\cos\theta_2 \right)\right]
     \,d\theta_{1}\,d\theta_{2}.
   \end{eqnarray}
Now in general we can consider  sampling from the joint probability density function

\begin{equation}
    h^{*}(\theta_1,\theta_2)=\frac{1}{C}\, h_{1}(\theta_1 \,|\, \theta_2)\, \left[\,h_{2}(\theta_2)\left(1+\frac{r}{R}\cos\theta_2 \right)\right], \mbox{~where~~} 0\leq \theta_1, \theta_2<2\pi
      \label{torus area dist}
\end{equation}
as a sampling scheme from the surface of a curved torus where the joint density on the parameter space or in the flat torus is pre-specified as $h(\theta_1, \theta_2).$ In particular, when $\theta_1$ and $\theta_2$ are independently distributed then $ h_{1}(\theta_1 \,|\, \theta_2)\propto h_{1}(\theta_1)$. Hence, the Eq. \ref{torus area dist} will reduce to
\begin{equation}
    h^{*}(\theta_1,\theta_2)\propto h_{1}(\theta_1)\, \left[h_{2}(\theta_2)\left(1+\frac{r}{R}\cos\theta_2 \right)\right], \mbox{~where~~} 0\leq \theta_1, \theta_2<2\pi.
    \label{torus independent dis}
\end{equation}
 We demonstrate the methodology in the following subsection to draw samples from different distributions on the surface of a curved torus maintaining the area proportionate dominating measure.

\subsection{Simulation Analysis} Here, we perform an extensive simulation study to draw samples  for different distributions on the surface of a curved torus.

\paragraph{\textbf{ (I) Uniform Distribution:}}
The simulation analysis in the Subsection-\ref{uniform simulation} is a special case of the methodology described in Section-\ref{torus sampling} where $\theta_1 $ and $\theta_2$ are independently uniformly distributed on the flat torus with respect to Lebesgue measure, and as a consequence, it will have the joint probability  density function $$h^{*}(\theta_1,\theta_2)=g(\theta_1,\theta_2)=\frac{1}{2\pi}\dfrac{(1+\frac{r}{R} \cos{\theta_{2}})}{2\pi},$$ from Eq. \ref{decompose equation}, Eq. \ref{torus independent dis},
 which implies $$ h_{1}(\theta_1)=\frac{1}{2\pi}, \mbox{~and~}
h_{2}(\theta_2)=\dfrac{(1+\frac{r}{R} \cos{\theta_{2}})}{2\pi}, \mbox{~where~~} 0\leq \theta_1, \theta_2<2\pi.$$

\begin{figure}[b]
	\includegraphics[width=\textwidth,height=0.55\textwidth]{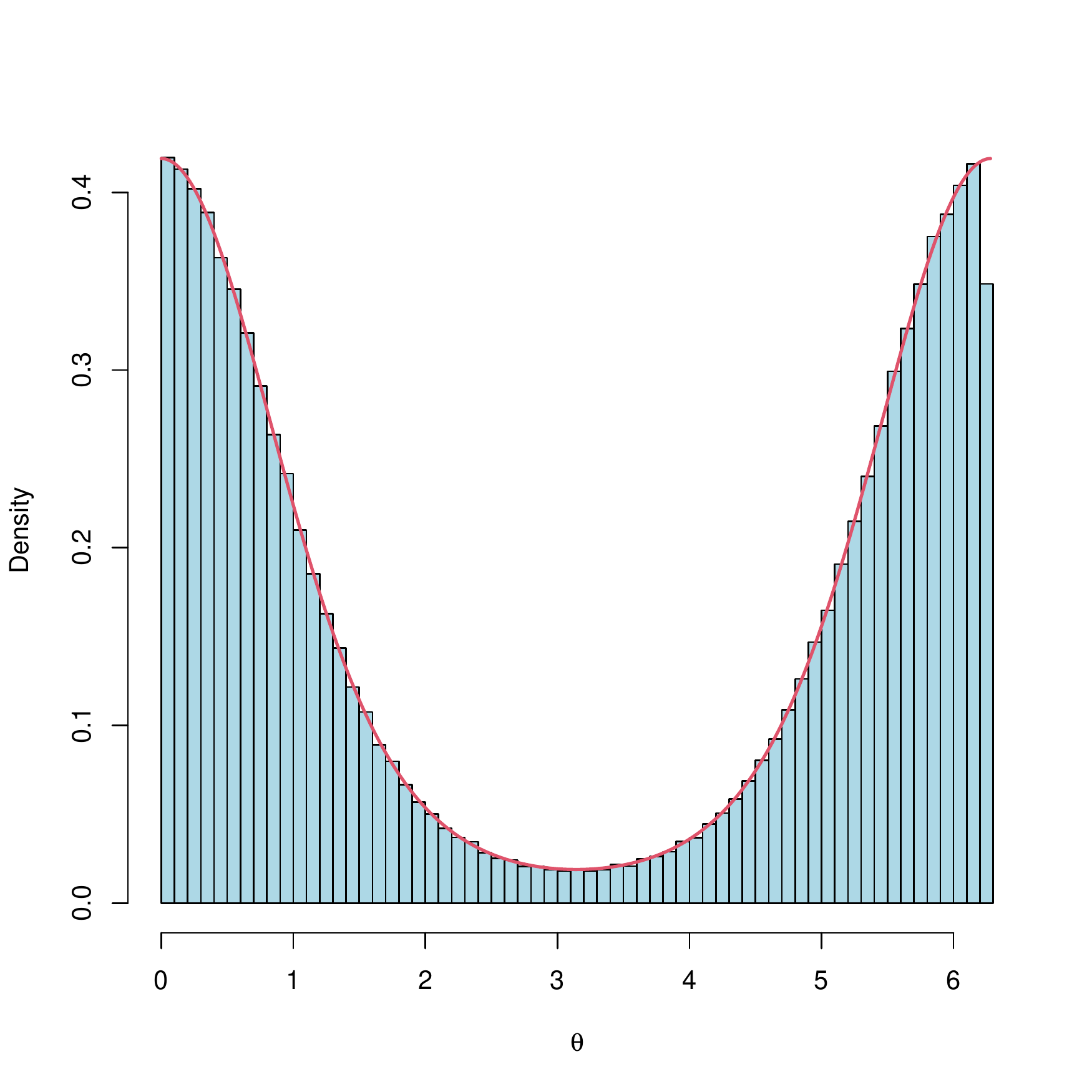}
	\caption{ Histogram of the data from the density  $\frac{e^{\cos\theta}~\left(1+\frac{r}{R}\cos\theta\right)}{2\pi\left[I_{0}(1)+\frac{r}{R}I_{1}(1)\right]}.$}
    \label{fig: density_cvm}
\end{figure}

\paragraph{\textbf{(II) von Mises Distribution:}}
To draw samples from the marginal von Mises distribution on the surface of the curved torus, let $\theta_1 $ and $\theta_2$ be independently von Mises  distributed on the flat torus with concentration parameters $\kappa_1$, $\kappa_2$, and location parameters $\mu_1$, $\mu_2$, respectively.  So, now it is necessary to draw samples from the target distribution

\begin{equation}
    h^{*}(\theta_1,\theta_2)= \frac{e^{\kappa_1\cos(\theta_1-\mu_1)}}{2\pi I_{0}(\kappa_1)} \left[ \frac{e^{\kappa_2\cos(\theta_2-\mu_2)}}{C} \left(1+\frac{r}{R}\cos\theta_2 \right)\right], 
    \label{von torus}
\end{equation}
where $0\leq \theta_1, \theta_2<2\pi$, $0\leq \mu_1,\mu_2<2\pi$, $\kappa_1,\kappa_2>0$, which implies 

$$ h_{1}(\theta_1)=\frac{e^{\kappa\cos(\theta_1-\mu_1)}}{2\pi I_{0}(\kappa_1)} \mbox{~and~}
 h_{2}(\theta_2)=\frac{e^{\kappa_2\cos(\theta_2-\mu_2)}}{C} \left(1+\frac{r}{R}\cos\theta_2 \right),$$
with $$C=2\pi\left[I_{0}(\kappa_2)+\frac{r}{R} \cos \mu_2~I_{1}(\kappa_2)\right]$$ is the normalizing constant (see, Appendix-\ref{appendix}). Hence, the joint probability density function in Eq. \ref{von torus} is the representative of the von Mises distribution on the surface of a curved torus.
Now let $\mu_1=\mu_2=0$, $\kappa_1=\kappa_2=1$, and $r=1.5,~R=3.$ then the below Figure-\ref{fig: density_cvm} is the histogram of the sampled data  from the distribution of the vertical angle $\theta_2$ using the HAR sampling method. 

\begin{figure}[t]
	\includegraphics[width=0.7\textwidth, height= 0.53\textwidth]{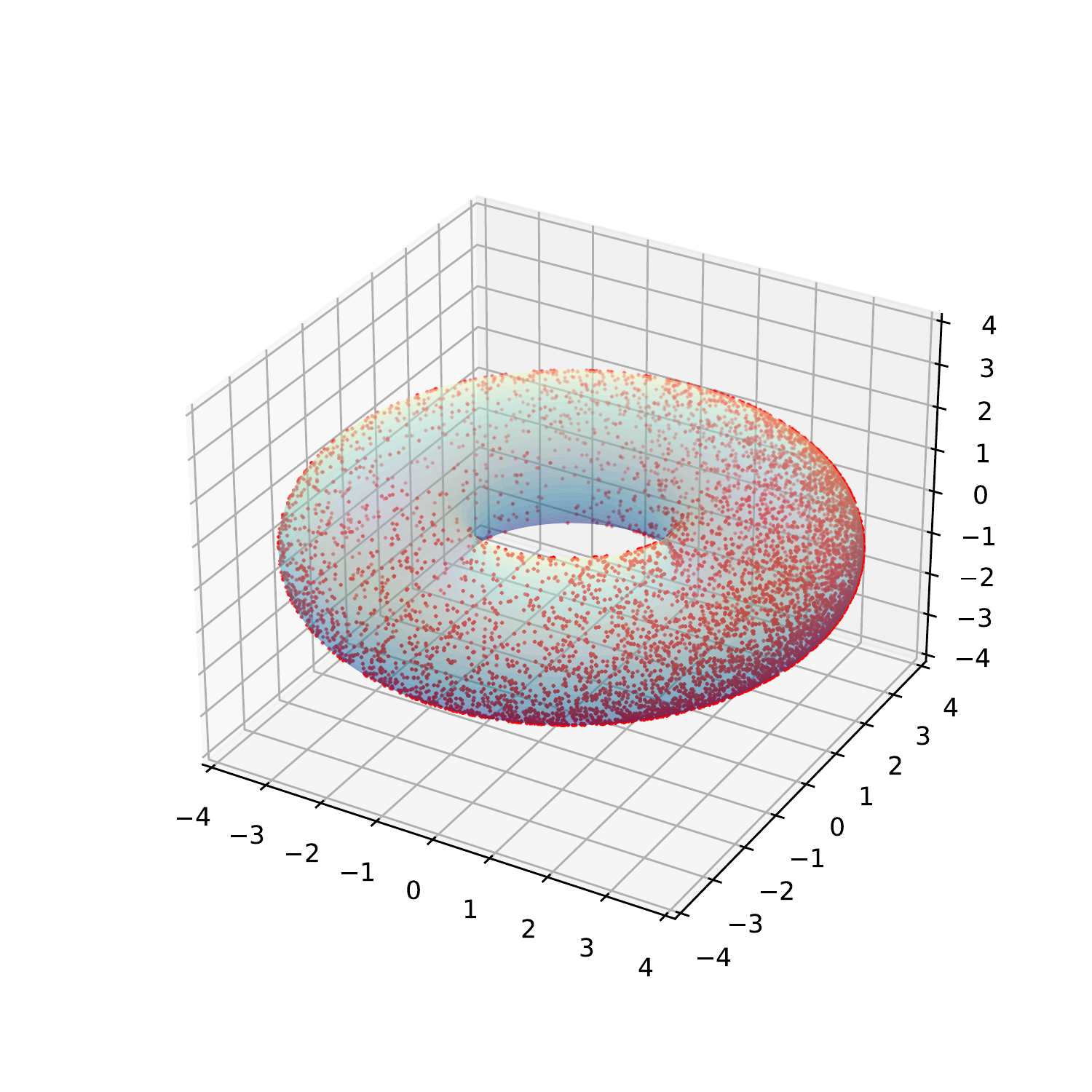}
	\caption{ Scatter plot of the  points on the torus using area measure for von Mises distribution.}
    \label{fig:torus vc area}
\end{figure}

\begin{figure}[t]
	\includegraphics[width=0.7\textwidth, height= 0.54\textwidth]{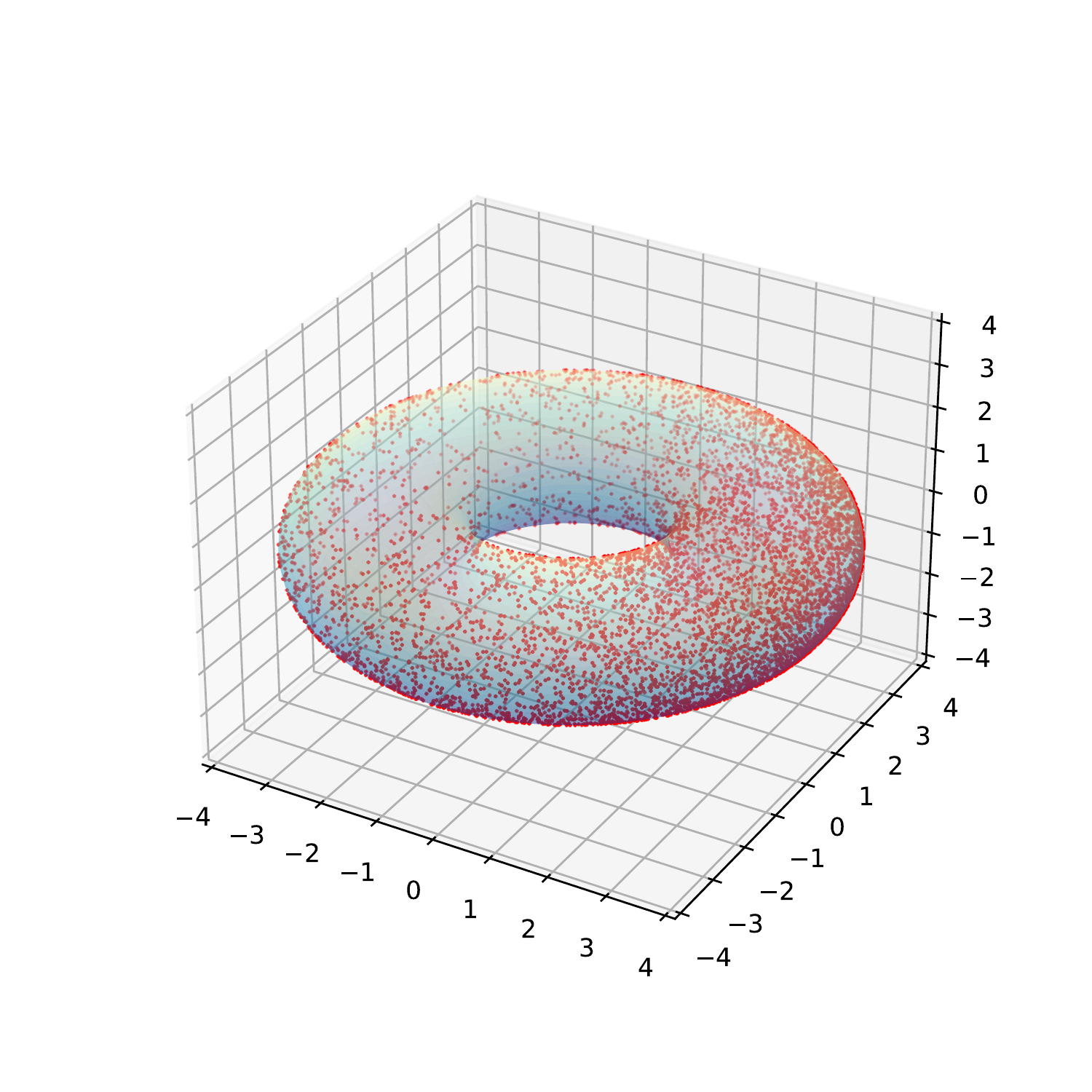}
	\caption{ Scatter plot of  points on torus when angular parameters are drawn from von Mises distribution from the flat torus.}
    \label{fig:torus vc para}
\end{figure}

Figures- \ref{fig:torus vc area} displays the scatter plot of sampled points from the von Mises distribution  from the surface of a curved torus with respect to area measure, and Figures-\ref{fig:torus vc para} displays the same when samples are drawn from von Mises on the flat torus and projected on the surface of a curved torus.

\newpage
\paragraph{\textbf{(III) Wrapped Cauchy Distribution:}}
The wrapped Cauchy distribution is one of the well-known circular distributions given by the probability density function in Eq. \ref{wrap cauchy}.

\begin{equation}
    f_{wc}(\theta)=\frac{1}{2\pi}\dfrac{1-\rho^2}{1+\rho^2-2\rho \cos({\theta-\mu})},
    \label{wrap cauchy}
\end{equation}
where $0\leq \theta<2\pi$, $0\leq \mu<2\pi$, and $0\leq \rho <1$.

Now we present the wrapped Cauchy distribution on the surface of the curved torus.  Let $\theta_1 $ and $\theta_2$ are independently wrapped Cauchy distributed with concentration parameter, $\rho_1,\rho_2$, and location parameter $\mu_1,\mu_2$, respectively on flat torus.  Therefore the corresponding joint density is given by

\begin{equation}
    h^{*}(\theta_1,\theta_2)=\dfrac{(2\pi)^{-1}(1-\rho_1^2)}{1+\rho_1^2-2\rho_1\cos({\theta_1-\mu_1})} \left[ \frac{(C~2\pi)^{-1}(1-\rho^2_2 )}{1+\rho_2^2-2\rho_2\cos({\theta_2-\mu_2})} \left(1+\frac{r}{R}\cos\theta_2 \right)\right], 
    \label{wc torus}
\end{equation}
where $0\leq \theta_1, \theta_2<2\pi$, $0\leq \mu_1,\mu_2<2\pi$, $0\leq \rho_1,\rho_2<1$, which implise

$$ h_{1}(\theta_1)= f_{wc}(\theta_1) \mbox{~and~} h_{2}(\theta_2)=\frac{1}{C}f_{wc}(\theta_2)\left(1+\frac{r}{R}\cos\theta_2 \right),$$
 with normalizing constant
$$C=\displaystyle \int_{0}^{2\pi} \left[ f_{wc}(\theta_2)~\left(1+\frac{r}{R}\cos\theta_2 \right) \right]~d\theta_2.$$

\begin{figure}[t]
	\includegraphics[width=.9\textwidth,height=.55\textwidth]{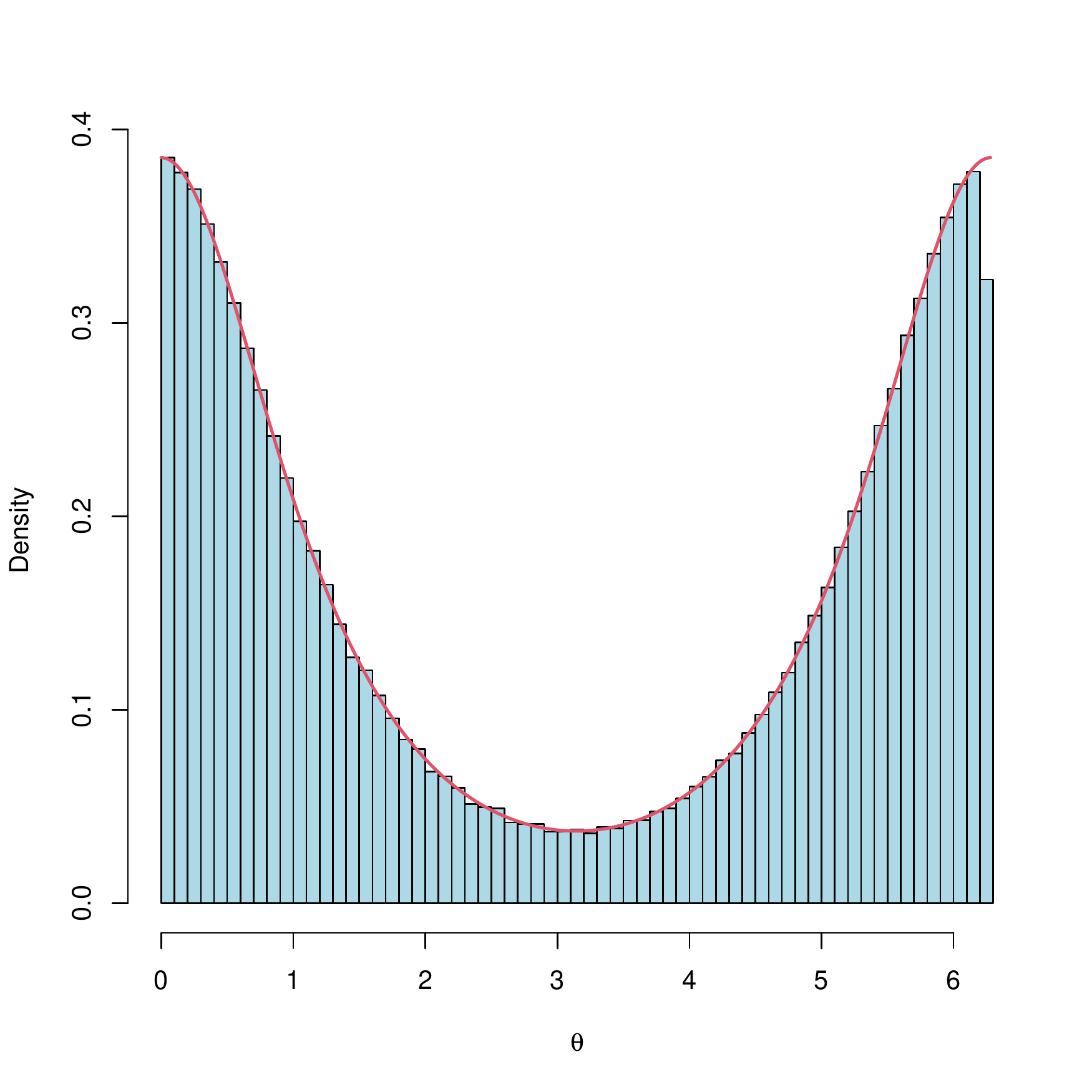}
	\caption{ Histogram of the data from the  density  $\frac{1}{C}f_{wc}(\theta)\left(1+\frac{r}{R}\cos\theta \right)$. }
    \label{fig:density_wc}
\end{figure}

Hence, the joint probability density function
in Eq. \ref{wc torus} is representative of the wrapped Cauchy distribution on the surface of a curved torus. Now, let us assume $\mu_1=\mu_2=0$, $\rho_1=\rho_2=0.3$, and $r=1.5,~R=3.$ then the below  Figure-\ref{fig:density_wc} is the histogram of the sampled data  from the distribution of the vertical angle $\theta_2$ using the HAR sampling method.

\begin{figure}[t]
	\includegraphics[width=0.7\textwidth, height= 0.53\textwidth]{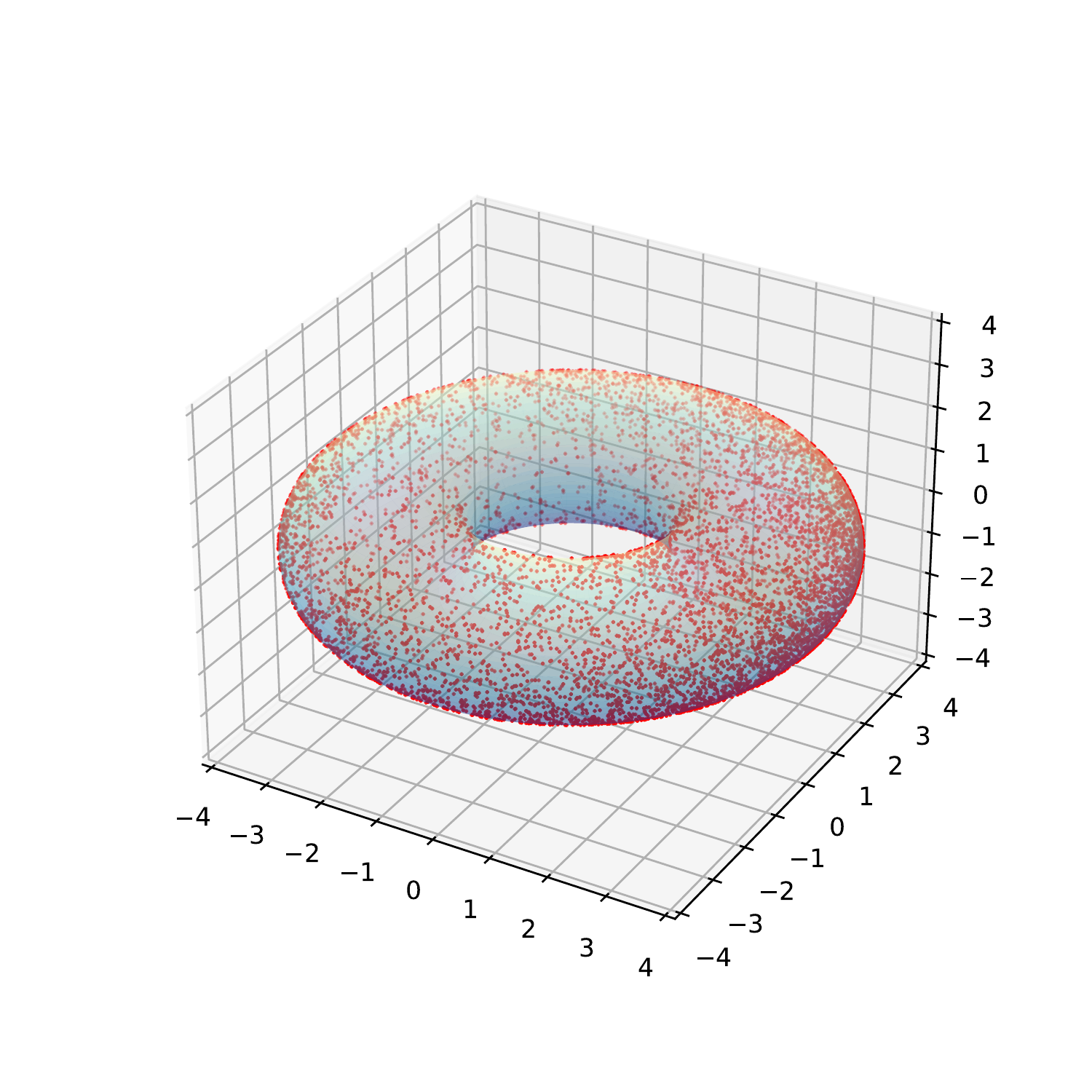}
	\caption{  Scatter plot of the  points on the torus using area measure for wrapped Cauchy distribution.}
    \label{fig: torus wc area}
\end{figure}

\begin{figure}[t]
	\includegraphics[width=0.7\textwidth, height= 0.53\textwidth]{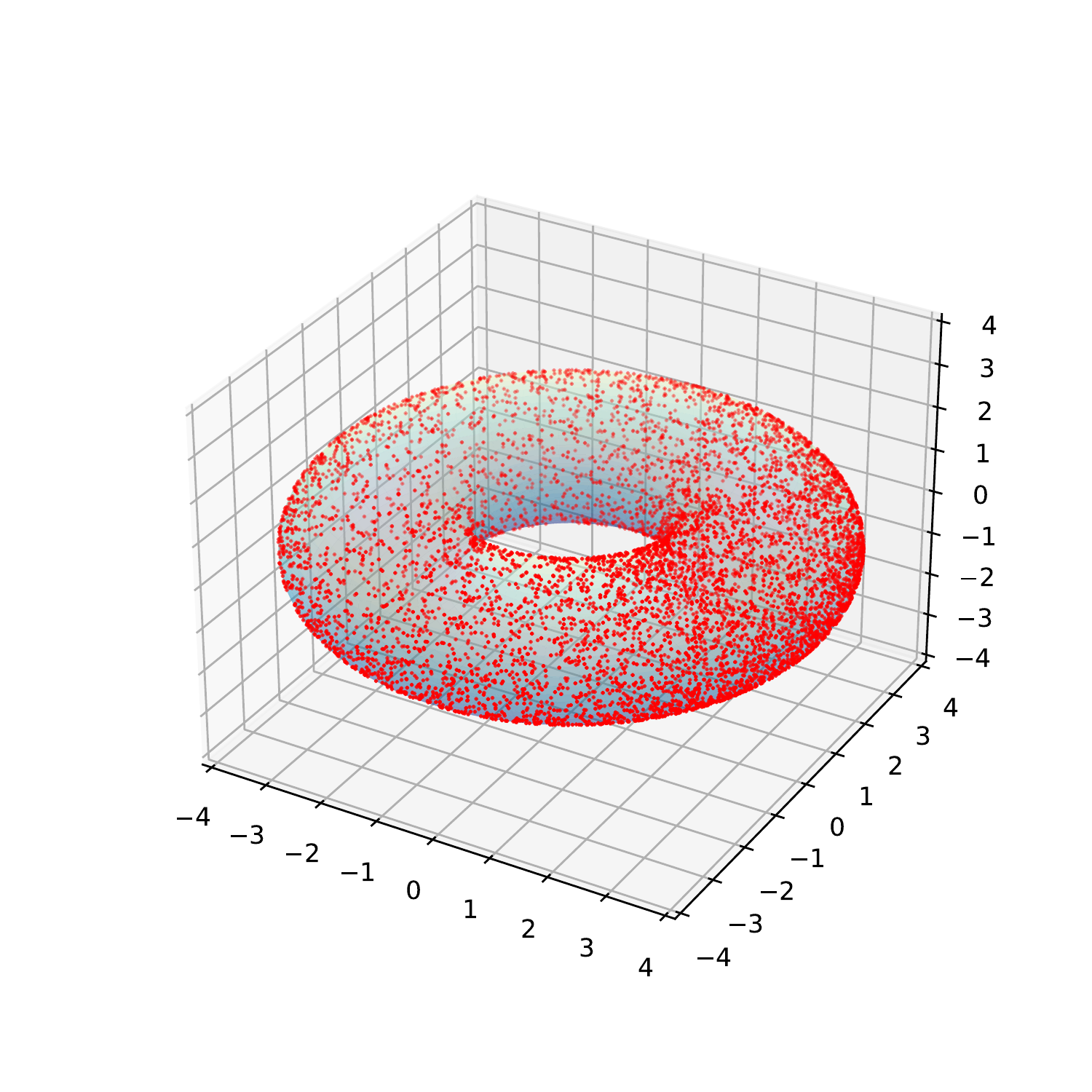}
		\caption{ Scatter plot of points on torus when angular parameters are drawn from wrapped Cauchy distribution from the flat torus.}
    \label{fig: torus wc para}
\end{figure}

Figures- \ref{fig: torus wc area} displays the scatter plot of sampled points from the wrapped Cauchy distribution  from the surface of a curved torus with respect to area measure, and Figures-\ref{fig: torus wc para} displays the same when samples are drawn from wrapped Cauchy on the flat torus and projected on the surface of a curved torus. 

\newpage
\paragraph{\textbf{ (IV) Kato and Jones Distribution:}}
The Kato and Jones distribution is a four-parameter family of circular distributions  that was first introduced by \cite{kato2010family} where they have used the M\"{o}bius transformation from $\Bar{\Theta}$ to $\Theta$ as

\begin{eqnarray}
 e^{i\Theta}&=& e^{i\mu}~\dfrac{e^{i\Bar{\Theta}}+\rho e^{i\nu}}{\rho e^{i(\Bar{\Theta}-\nu)}+1}  \hspace{1.4cm}\mbox{or}     \nonumber\\
    \Theta &=& \mu+\nu+2\arctan\left[ \left(\dfrac{1-\rho}{1+\rho} \right) \tan \left(\frac{\Bar{\Theta}-\nu}{2}\right)\right],
    \label{mobius}
\end{eqnarray}

where  $0\leq \mu,\nu<2\pi$, and $0\leq \rho <1$.

Now, they have taken the random variable $\Bar{\Theta}$ from von Mises distribution with $\mu=0$ and $\kappa$, and then using the transformation in Eq. \ref{mobius} they get the desired density as

\begin{equation}
    f_{kj}(\theta)= \frac{(1-\rho^2)(2\pi I_{0}(k))^{-1}}{1+\rho^2-2\rho \cos{(\theta-\gamma)}} \exp\left[ \dfrac{\kappa \{ \xi \cos{(\theta-\eta)}-2\rho \cos \nu \}}{1+\rho^2-2\rho \cos{(\theta-\gamma)}}   \right],
    \label{kato jons}
\end{equation}

where $0\leq \mu,\nu<2\pi$, and $0\leq \rho <1$, $\kappa>0$, and 
$\gamma=\mu+\nu$, $\xi=\sqrt{\rho^4+2\rho^2\cos{(2\nu)}+1}$, $\eta=\mu+\arg(\rho^2\cos{(2\nu)}+1+i\rho^2\sin{(2\nu)})$\

\begin{figure}[t]
\includegraphics[width=0.9\textwidth,height=0.55\textwidth]{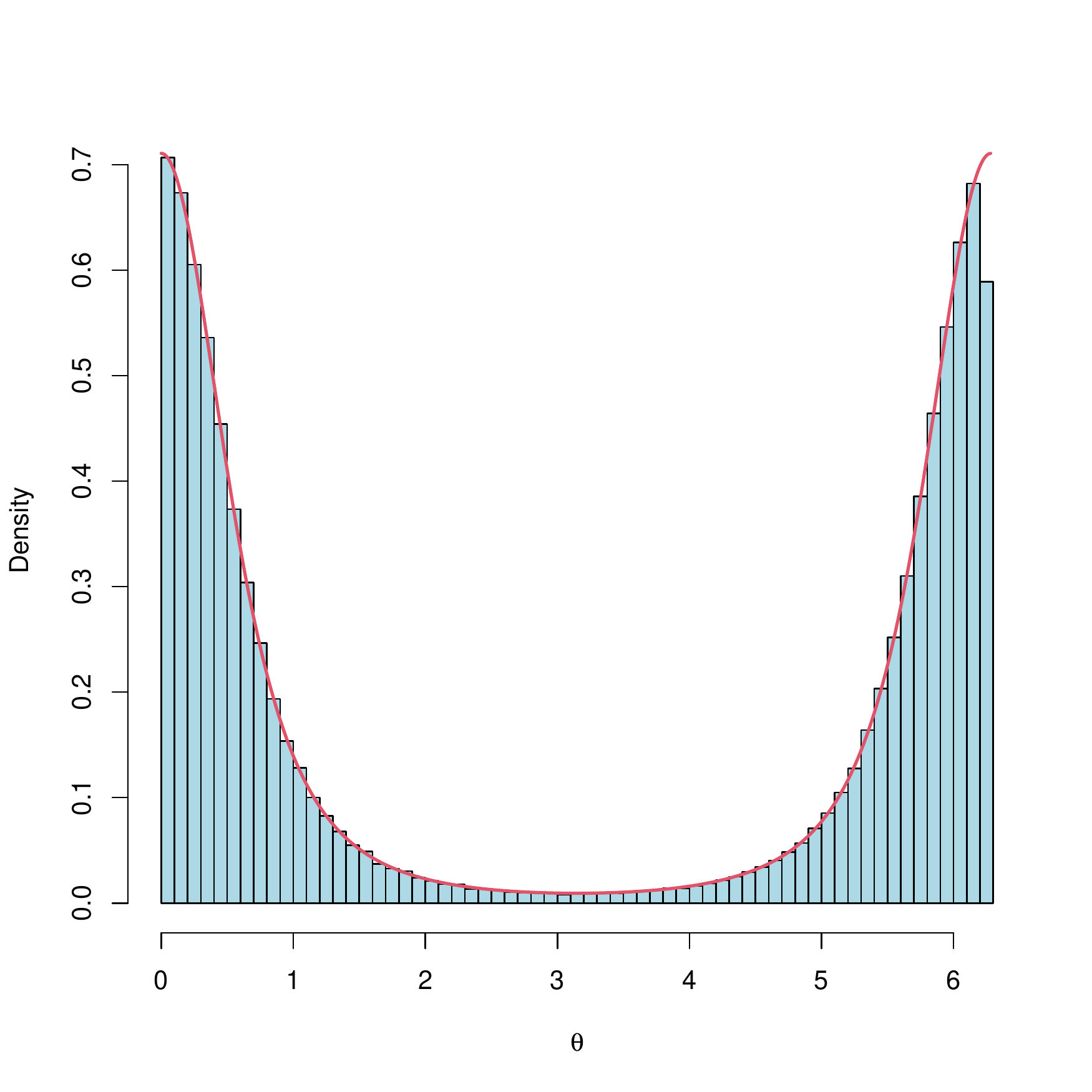}
	\caption{ Histogram of the data from the  density  $\frac{1}{C}f_{kj}(\theta)\left(1+\frac{r}{R}\cos\theta \right).$}
    \label{fig:density_kj}
\end{figure}

\begin{figure}[b]
	\includegraphics[width=0.7\textwidth, height= 0.53\textwidth]{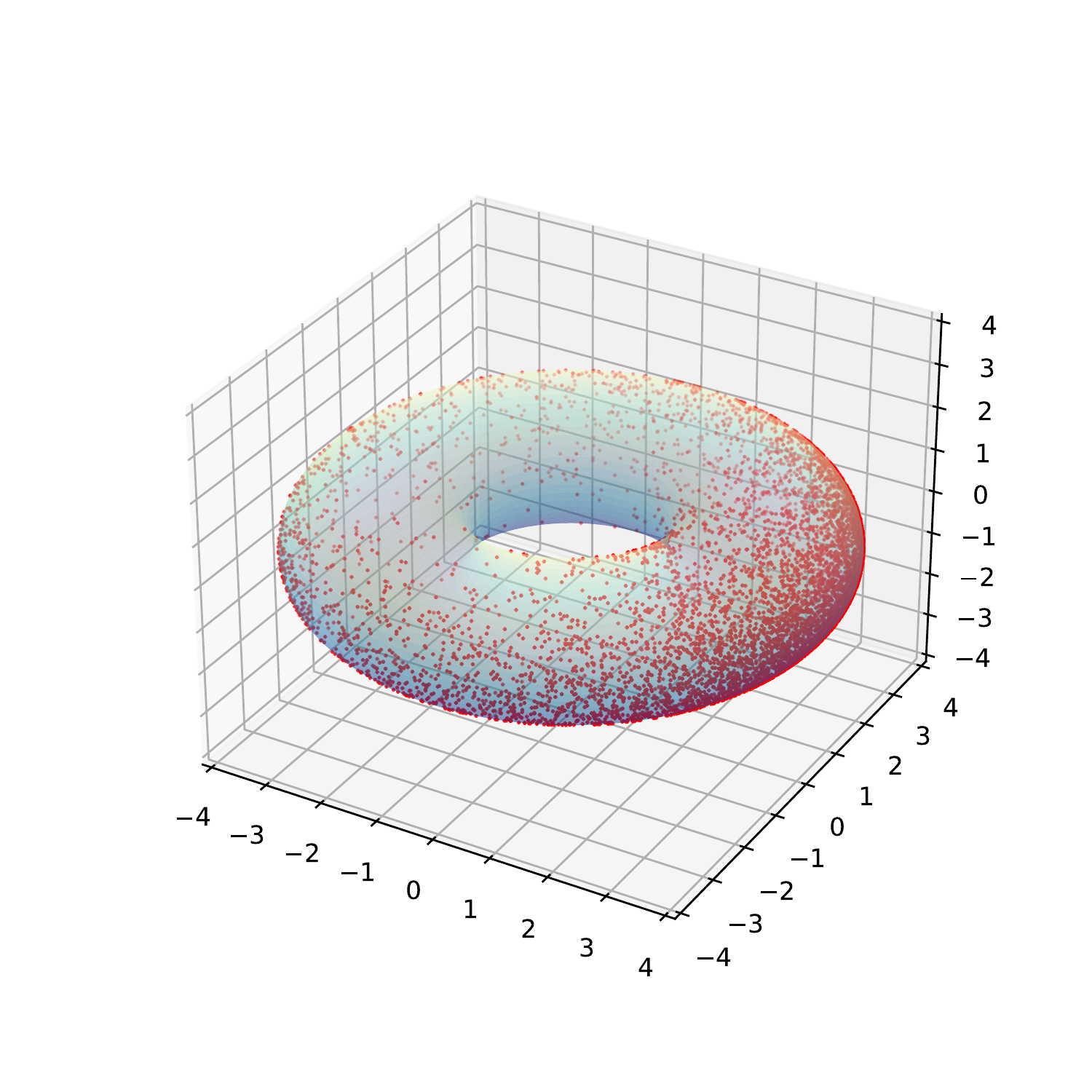}
	\caption{Scatter plot of the  points on the torus using area measure for Kato and Jones distribution.}
    \label{fig: torus kj area}
\end{figure}

\begin{figure}[t]
	\includegraphics[width=0.7\textwidth, height= 0.53\textwidth]{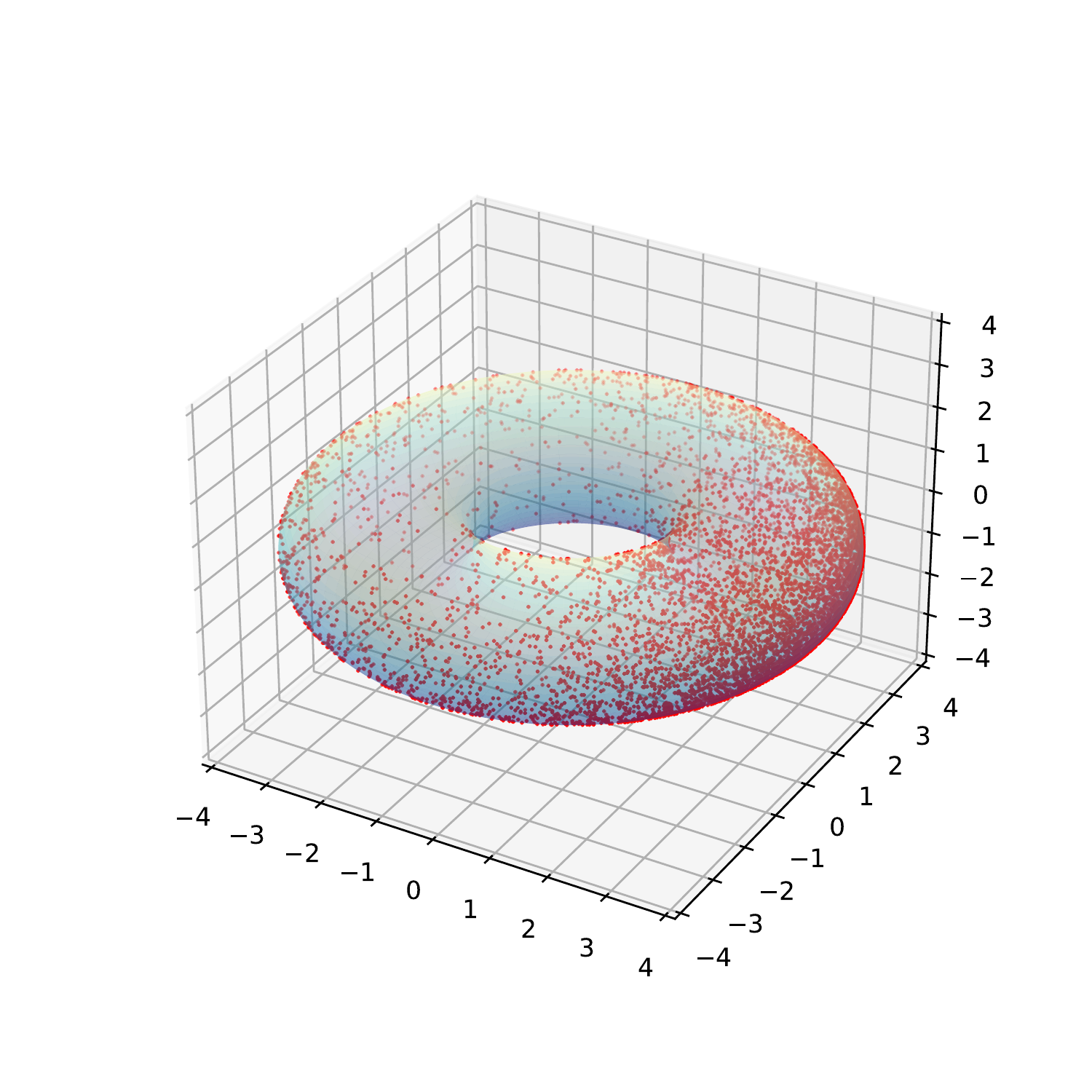}
		\caption{ Scatter plot of points on torus when angular parameters are drawn from Kato and Jones distribution from the flat torus.}
    \label{fig: torus kj para}
\end{figure}

Now we present the Kato and Jones  distribution  on the surface of the curved torus.  d Jones  distribution on the surface of the curved torus.  Let $\theta_1 $ and $\theta_2$ are independently Kato and Jones distributed with concentration parameters, $\rho_1,\rho_2,\kappa_1, \kappa_2$, and location parameter $\mu_1,\mu_2,\nu_1,\nu_2$, respectively on flat torus.  Therefore the corresponding joint density is given by
\begin{multline}
    h^{*}(\theta_1,\theta_2)= \frac{(1-\rho_1^2)(2\pi I_{0}(\kappa_1))^{-1}}{1+\rho_1^2-2\rho_1\cos{(\theta_1-\gamma_1)}} \exp\left[ \dfrac{\kappa_1 \{ \xi_1 \cos{(\theta_1-\eta_1)}-2\rho_1 \cos \nu_1 \}}{1+\rho_1^2-2\rho_1 \cos{(\theta_1-\gamma_1)}}  \right] \times\\ 
     \left[\frac{ (1-\rho_2^2)(C~2\pi I_{0}(\kappa_2))^{-1} \left (1+\frac{r}{R}\cos\theta_2 \right)}{1+\rho_2^2-2\rho_2\cos{(\theta_2-\gamma_2)}}    \exp\left[ \dfrac{\kappa_2 \{ \xi_2 \cos{(\theta_2-\eta_2)}-2\rho_2 \cos \nu_2 \}}{1+\rho_2^2-2\rho_2 \cos{(\theta_2-\gamma_2)}}\right]  \right],
    \label{kj torus}
\end{multline}
where $0\leq \mu_j,\nu_j<2\pi$, and $0\leq \rho_j <1$, $\kappa_j>0$, and 
$\gamma_j=\mu_j+\nu_j$, $\xi_j=\sqrt{\rho_j^4+2\rho_j^2\cos{(2\nu_j)}+1}$, $\eta_j=\mu_j+\arg(\rho_j^2\cos{(2\nu_j)}+1+i\rho_j^2\sin{(2\nu_j)}), \mbox{~for~} j=1,2,$
which implies that

$$ h_{1}(\theta_1)= f_{kj}(\theta_1) \mbox{~and~} h_{2}(\theta_2)=\frac{1}{C}f_{kj}(\theta_2)\left(1+\frac{r}{R}\cos\theta_2 \right),$$
 with normalizing constant
$$C=\displaystyle \int_{0}^{2\pi} \left[ f_{kj}(\theta_2)~\left(1+\frac{r}{R}\cos\theta_2 \right) \right]~d\theta_2.$$
Hence, the joint probability density function
in Eq. \ref{kj torus} is representative of the Kato and Jones distribution on the surface of a curved torus.
Now let us consider $\mu_1=\mu_2=\nu_1=\nu_2=0$, $\rho_1=\rho_2=0.3$, $\kappa_1= \kappa_2=1$ , and $r=1.5,~R=3.$ then the below  Figure-\ref{fig:density_kj} is the histogram of the sampled data  from the distribution of the vertical angle $\theta_2$ using the HAR sampling method.
Figures- \ref{fig: torus kj area} displays the scatter plot of sampled points from the von Mises distribution  from the surface of a curved torus with respect to area measure, and Figures-\ref{fig: torus kj para} displays the same when samples are drawn from von Mises on the flat torus and projected on the surface of a curved torus. 

\newpage
\paragraph{\textbf{Remark:}}
The  Algorithm-\ref{alg:algo HAR} can be more efficiently implemented for batch sampling in a large number desired density function $f(x)$ with support $[a,b]$ satisfying the necessary conditions. To implement this, we select the left-most partition, draw the required number of random samples from the envelope density for that specific cell, and apply the acceptance-rejection method. The accepted samples are retained in that cell, while the rejected samples are forwarded to the next cell. Now in the next cell, we calculate the required number of random samples and subtract it from the samples already present. We then generate the number of random samples from the envelope density, applying the acceptance-rejection method again. The accepted samples remain in that cell, and the rejected samples are moved to the next cell. We repeat this process for all the partitions from left to right, followed by random permutation of the generated samples, providing the random samples in batches from the desired distribution. 
Table-\ref{table:vm HAR batch1}, and Table-\ref{table:vm HAR batch2} compare the acceptance percentages of sample sizes for the modified HAR method of batch sampling and the existing vMBFR method for the von Mises distribution.

\begin{table}[h!]
\centering
\begin{tabular}{|l|c|c|c|c|c|c|c|c|c|c|}
\hline
$\kappa$ & $0.1$ & $0.2$ & $0.3$ & $0.4$ & $0.5$ & $0.6$ & $0.7$ & $0.8$ & $0.9$ & $1$\\
\hline
\hline
\textbf{HAR } & $99.76$ & $99.76$ & $99.73$ & $99.75$ & $99.72$ & $99.74$ & $99.81$ & $99.77$ & $99.73$ & $99.79$\\

\hline
\textbf{vMBFR } & $99.76$ & $99.06$  & $97.90$ & $96.67$ & $95.04$   & $93.23$  & $91.88$  & $89.88$ & $88.12$ & $86.94 $\\
\hline
\end{tabular}
\vspace{.1cm}
\caption{Acceptance percentage comparison table for von Mises distribution for batch .}
\label{table:vm HAR batch1}

\centering
\scalebox{1}{
\begin{tabular}{|l|c|c|c|c|c|c|c|c|c|c|}
\hline
$\kappa$ & $2$ & $3$ & $4$ & $5$ & $10$ & $20$ &$40$ & $60$  &$80$ &$100$ \\
\hline
\hline
\textbf{HAR } & $99.87$ & $99.81$ & $99.87$ & $99.72$ & $99.68$ & $99.67$ &$99.85$ & $99.93$ &$99.89$ & $99.92$ \\

\hline
\textbf{vMBFR } & $76.95$ & $72.37$  & $69.96$ & $69.46$ & $67.46$   & $66.64$  & $66.43$  & $65.96$ &$65.94$ &$65.69$\\
\hline
\end{tabular}}
\vspace{.1cm}
\caption{Acceptance percentage comparison table for von Mises distribution for batch.}
\label{table:vm HAR batch2}
\end{table}

The Algorithm-\ref{alg:algo HAR batch} provides the pseudo-code for the proposed modified HAR method for batch sampling.

\SetKwComment{Comment}{/* }{ */}
\RestyleAlgo{ruled}
\begin{algorithm}
\caption{HAR Batch sampling algorithm}\label{alg:algo HAR batch}
\KwData{ Target probability density function $f(x)$ with support $[a,b]$.}
$na \gets$ Number of random samples required\;
$np \gets$ Number of partitions with equal length\;
$pt\gets$  A sequence in $[a,b]$ with length $(np+1)$ \;
$H\gets f(pt)$ \Comment*[r]{ Heights of the probability density function at pt}
$H_m\gets$ A sequence of the maximum heights of each portion\;
$bl\gets \frac{b-a}{np}$ \Comment*[r]{ Bin length } 
$A\gets \displaystyle \sum H_m*bl$ \Comment*[r]{ Total Area which is larger than one. }
$p_m\gets \displaystyle \frac{H_m}{\sum H_m}$\;
$n\gets   \left\lfloor na*A\right\rfloor+np$ \Comment*[r]{Proposed number of random samples.}
$ng \gets 0 $\Comment*[r]{Initializing the count for number of  sample generated.}
$y_0 \gets \mbox{~iniciated as array of size zero} $\;
$y \gets \mbox{~iniciated as array of size zero} $\;
\For{$i \gets 1 \mbox{~ to~} np$}{
  $n_i \gets \max\{ \lfloor n*p_m[i] \rfloor -\mbox{~ length}(y_0),0\}$\;
  \eIf{$n_i>0$ }{
    $d_i\gets$ draw $ n_i$ number random samples from $U[0,1]$ \;
  }{
  $d_i\gets $ array of size zero\;
  }
  $ng \gets ng+d_i $\;
  $x_i\gets \mbox{marge}(pt[i]+(d_i*bl),(y_0+bl)) $\;
  $px\gets \frac{f(x_i)}{H_m[i]}$\;
  $rp \gets$ For each of $px$ draw a random sample from respective $Bernoulli(px)$\;
  \eIf{$rp=1$}{
    $y_1\gets$  $x_i$\;
  }{\If{$rp=0$}{
  $y_0\gets$  $x_i$\;
  }
  }
  $y\gets \mbox{marge} (y,y_1)$ \Comment*[r]{Desires random sample upto ith cell.}
}
$y\gets$ a ramdom permutation of $y$\;
\KwResult{Samples from the desired probability density function $f(x)$ with support $[a,b]$.}
\end{algorithm}

\section{Conclusion}
To understand  any  distribution on the surface of the curved  torus, it is necessary  to have a uniform random sample from there. In this article, we introduce  a probabilistic transformation that enables us  to draw random samples from the uniform  distribution on the surface of a curved  torus  without any rejection of the data. The proposed EAU sampling method provides a substantial improvement over the acceptance-rejection sampling  in this context.   We also introduce a new genesis of random samples from some popular circular distributions using  HAR sampling. Converting the step function, which provides an upper-Riemann-sum of an integral, into a legitimate probability density function,  we constructed a very thin envelope to the target density on a circle. We synthesize the equivalent sampling scheme from the surface area of a curved torus for which the marginal densities are  pre-specified  on the flat torus. Finally,  the idea has been  generalized  to draw random samples from different distributions on the surface of the curved torus incorporating its intrinsic geometry with a high acceptance rate. The proposed methods of the different genesis of data from the circle and the surface of a curved torus not only add computational advantage to the probabilistic investigations but also speed up the simulated experiments in statistical inferences.

\newpage
\section{Appendix} \label{appendix}
\subsection{A1}\label{appendix eau}
Here we provide the proof of Theorem \ref{EAU thm}.
\begin{proof}

   \textbf{Part-I:}
    In this case, we consider $U<p(x)$, and for $X>\pi$ or $X<\pi$ we have $Y=X.$ Hence, we have
    
$$
P(Y \leq y)= \int_{0}^{y} \frac{1}{2\pi} P(U<p(x))  \,dx 
$$
$$\hspace{2cm}=\frac{1}{2\pi} \int_{0}^{y} (1+a\cos x) \,dx .$$ Therefore, the integral gives
$$
P(Y \leq y)= \frac{1}{4\pi} \left(y+a \sin{y}\right)
$$

 \textbf{Part-II:} 
    In this case, we consider $U>p(x)$. Hence, we have
    \begin{equation*}
Y = \left\{
        \begin{array}{ll}
            \pi-X & \text{when} \quad X<\pi\\
            3\pi-X & \text{when} \quad X>\pi
        \end{array}
    \right.
\end{equation*}
Now, when $0<X<\pi$ then $0<Y<\pi$. So, we have 

$$
P(Y \leq y)= \int_{\pi-y}^{\pi} \frac{1}{2\pi} P(U>p(x))  \,dx 
$$
$$\hspace{2.3cm}=\frac{1}{4\pi} \int_{\pi-y}^{\pi} (1-a\cos x) \,dx.$$
Therefore, the integral gives
$$
P(Y \leq y)= \frac{1}{4\pi} \left(y+a \sin{y}\right)
.$$

Now, when $\pi<X<2\pi$ then $\pi<Y<2\pi$. So, we have 

$$
P(Y \leq y)= \int_{0}^{\pi} \frac{1}{2\pi} P(U>p(x)) \,dx  + \int_{3\pi-y}^{2\pi} \frac{1}{2\pi} P(U>p(x))  \,dx 
$$

$$\hspace{2.3cm}=\frac{1}{4\pi} \int_{0}^{\pi} (1-a\cos x) \,dx + \frac{1}{4\pi} \int_{3\pi-y}^{2\pi} (1-a\cos x) \,dx$$
Therefore, the integral gives
$$
P(Y \leq y)= \frac{1}{4\pi} \left(y+a \sin{y}\right)
.$$

Adding the two probabilities in \textbf{Part-I} and \textbf{Part-II} we get
$$
P(Y \leq y)= \frac{1}{2\pi} \left(y+a \sin{y}\right)
.$$
Hence, $G(y)=\frac{1}{2\pi} \left(y+a \sin{y}\right)$, and the theorem follows.
\end{proof}

\subsection{A2} \label{appendix acceptence}

\begingroup
\allowdisplaybreaks
\begin{align}
 \frac{1}{M} \int_{a}^{x} f(t)\,\,dt 
     &=  \sum_{i=1}^{k} \left[ \frac{1}{M} \int_{A_i} \textbf{I}_{(t\leq x)}\,\, \dfrac{f(t)}{p(t)} \,\, p(t) \,\, dt \right] \nonumber\\
     &=  \sum_{i=1}^{k} \left[ \frac{H_{i}}{M} \int_{A_i} \textbf{I}_{(t\leq x)}\,\, \dfrac{f(t)}{H_{i}\,p(t)} \,\, p(t) \,\, dt \right] \nonumber\\
     &=  \sum_{i=1}^{k} \left[ \frac{H_{i}}{M} \int_{A_i} \textbf{I}_{(t\leq x)}\,\, \dfrac{f(t)/P(A_{i})}{H_{i} /P(A_{i})} \,\,  \left(\frac{ \displaystyle \int_{A_{i}}p(t)\,dt}{p(t)}\right)\, \left(\frac{p(t)}{\displaystyle \int_{A_{i}}p(t)\,dt}\right) \,\, dt \right] \nonumber\\
     &=  \sum_{i=1}^{k} \left[ \frac{B\,H_{i}}{M} \int_{A_i} \textbf{I}_{(t\leq x)}\,\, \dfrac{f(t)/P(A_{i})}{B\,H_{i} /P(A_{i})} \,\,  \left(\frac{ 1}{\frac{1}{B}}\right)\, \left(\frac{1}{B}\right) \,\, dt \right] \nonumber\\
     &=  \sum_{i=1}^{k} \left[ \frac{B\,H_{i}}{M} \int_{A_i} \textbf{I}_{(t\leq x)}\,\, \dfrac{f(t)/P(A_{i})}{M_{i}\, \left(\frac{1}{B}\right)} \,\, \left(\frac{1}{B}\right) \,\, dt \right] \nonumber\\
     &=  \sum_{i=1}^{k} \left[ \frac{B\,H_{i}}{\displaystyle B\, \sum_{i=1}^{k}H_{i}} \int_{A_i} \textbf{I}_{(t\leq x)}\,\, \dfrac{f(t)/P(A_{i})}{M_{i}\, \left(\frac{1}{B}\right)} \,\, \left(\frac{1}{B}\right) \,\, dt \right] \nonumber\\
     &=  \sum_{i=1}^{k} \left[ \frac{H_{i}}{\displaystyle  \sum_{i=1}^{k}H_{i}} \frac{1}{M_{i}} \left( \frac{1}{1/M_{i}}  \right)
     \int_{A_i} \textbf{I}_{(t\leq x)}\,\, \dfrac{f(t)/P(A_{i})}{M_{i}\, \left(\frac{1}{B}\right)} \,\, \left(\frac{1}{B}\right) \,\, dt \right] \nonumber\\
     &=  \sum_{i=1}^{k} \left[ \frac{H_{i}}{\displaystyle  \sum_{i=1}^{k}H_{i}} \frac{P(A_{i})}{B\,H_{i}} \left( \frac{1}{1/M_{i}}  \right)
     \int_{A_i} \textbf{I}_{(t\leq x)}\,\, \dfrac{f(t)/P(A_{i})}{M_{i}\, \left(\frac{1}{B}\right)} \,\, \left(\frac{1}{B}\right) \,\, dt \right] \nonumber\\
     &=  \sum_{i=1}^{k} \left(\frac{P(A_{i})}{M} \right) \left[ 
     \dfrac{\displaystyle \int_{A_i} \textbf{I}_{(t\leq x)}\,\, P\left( U<\dfrac{f(t)/P(A_i)}{M_i(\frac{1}
{B})}\right) \,\, \left(\frac{1}{B}\right) \,\, dt }{\displaystyle \frac{1}{1/M_{i}} }
     \right] \nonumber\\
     \nonumber\\
     \label{appendix rejection sampling21}
\end{align}%
\endgroup

\subsection{A3} \label{appendix har}

 We will determine the normalizing constant C for the von Mises distribution on a torus' surface. We will utilize the subsequent two Bessel function identities to achieve this objective.

\begin{equation}
    \frac{1}{2\pi}\int_{0}^{2\pi} e^{\kappa\cos{\theta}} \cos{p\theta}~d\theta=I_p(\kappa),
    \label{i0k}
\end{equation}
and
\begin{equation}
   \frac{1}{2\pi}\int_{0}^{2\pi} e^{\kappa\cos{\theta}} \sin{n\theta}~d\theta=0.
   \label{ink}
\end{equation}

Now, the constant $C$ can be written as
\begin{eqnarray}
    C&=& \displaystyle \int_{0}^{2\pi} \left[e^{\kappa\cos(\theta-\mu)}\left(1+\frac{r}{R}\cos\theta \right)\right]~d\theta \nonumber\\
     &=&\displaystyle \int_{0}^{2\pi} e^{\kappa\cos(\theta-\mu)}~d\theta + \int_{0}^{2\pi} \left(1+\frac{r}{R}\cos\theta \right)~d\theta \nonumber\\
     &=& C_1+C_2,
     \label{c}
\end{eqnarray}
where using Eq. \ref{i0k} we get
\begin{equation}
    C_1=\displaystyle \int_{0}^{2\pi} e^{\kappa\cos(\theta-\mu)}~d\theta=2\pi I_0(\kappa),
    \label{c1}
\end{equation}
 and 

$$C_2=\frac{r}{R}\int_{0}^{2\pi} e^{\kappa\cos{(\theta-\mu)}}\cos\theta ~d\theta .$$

Now for $C_2$, let $\theta-\mu=t$ $\implies~ d\theta=dt, \mbox{~and ~}\theta=\mu+t.$ Therefore, $C_2$ becomes
\begin{eqnarray}
    C_2&=& \frac{r}{R}\int_{-\mu}^{2\pi-\mu} e^{\kappa\cos t}\cos {(\mu+t)} ~dt  \nonumber\\
    &=&\frac{r}{R}\int_{-\mu}^{2\pi-\mu} e^{\kappa\cos t}\left( \cos t \cos \mu -\sin t \sin \mu \right)~dt \nonumber\\
    &=&  \frac{r \cos \mu}{R}\int_{-\mu}^{2\pi-\mu} e^{\kappa\cos t} \cos t ~dt +   \frac{r \sin \mu}{R}\int_{-\mu}^{2\pi-\mu} e^{\kappa\cos t} \sin t ~dt \nonumber
\end{eqnarray}

Now, using the identities from Eq. \ref{i0k}, and Eq. \ref{ink} we can write
\begin{equation}
    C_2=\frac{r }{R} \cos \mu ~2\pi I_1(\kappa).
    \label{c2}
\end{equation}

Substituting the values from the Eq. \ref{c1}, and Eq. \ref{c2} in the Eq. \ref{c} we get
 \begin{equation*}
     C=2\pi \left[I_0(\kappa)+ \frac{r }{R} \cos \mu ~I_1(\kappa) \right].
     \label{ccom}
 \end{equation*}

\newpage

\begin{scriptsize}
\bibliographystyle{natbib}
	\bibliography{buddha_bib.bib}
\end{scriptsize}

\end{document}